\documentclass[journal, 12pt, draftclsnofoot, onecolumn]{IEEEtran}
\usepackage{cite}
\usepackage{balance}
\usepackage{color}
\usepackage{epsfig}
\usepackage{epstopdf}
\usepackage{graphicx}
\usepackage{tabularx}
\usepackage{booktabs}

\usepackage{varwidth}% http://ctan.org/pkg/varwidth
\usepackage{balance}

\usepackage{moreverb,algorithmic}
\usepackage{amsmath}
\usepackage{tabularx}
\usepackage[ruled]{algorithm2e}

\usepackage{amssymb,amsmath,amsthm}
\newcommand{\maestro}{{${\tt Maestro}$}}
\newcommand{\dedup}{{${\tt Dedup}$}}

\begin{document}
\title{Robust Orchestration of Concurrent Application Workflows in Mobile Device Clouds}
\author{Parul~Pandey,~\IEEEmembership{Student Member,~IEEE,} Hariharasudhan~Viswanathan,~\IEEEmembership{Student Member,~IEEE,}
        and~Dario~Pompili,~\IEEEmembership{Senior Member,~IEEE}% <-this % stops a space
\thanks{The authors are with the Department of Electrical and Computer Engineering, Rutgers University, New Brunswick, NJ. Their emails are \{parul\_pandey, hari\_viswanathan, pompili\}@cac.rutgers.edu. A preliminary shorter version of this work appeared in the Proc. of Workshop on Distributed Adaptive Systems~(DAS) at the IEEE Intl. Conference on Autonomic Computing~(ICAC), July'16~\cite{viswanathan2016maestro}. This work was supported by the Office of Naval Research---Young Investigator Program~(ONR-YIP) Grant No.~11028418.}}
\maketitle
%\author{Parul~Pandey,~\IEEEmembership{Student~Member,~IEEE,}
%        Hariharasudhan~Viswanathan,~\IEEEmembership{Student~Member,~IEEE,}
%        and~Dario~Pompili,~\IEEEmembership{Senior Member,~IEEE}% <-this % stops a space
%\IEEEcompsocitemizethanks{\IEEEcompsocthanksitem P. Pandey, H. Viswanathan, and D. Pompili are with the Dept. of Electrical and Computer Engineering, Rutgers University--New Brunswick, NJ, USA.\protect\\
%E-mails: \{parul\_pandey, hari\_viswanathan, pompili\}@cac.rutgers.edu\protect\\
%\IEEEcompsocthanksitem A preliminary shorter version appeared in the Proc. of Workshop on Distributed Adaptive Systems~(DAS) at the IEEE Intl. Conference on Autonomic Computing~(ICAC), July'16~\cite{viswanathan2016maestro}.\protect\\
%\IEEEcompsocthanksitem This work was supported in part by the Office of Naval Research---Young Investigator Program~(ONR-YIP) Grant No.~11028418.
%}
%\thanks{}
%}

%\markboth{Submitted to IEEE Transactions on Parallel and Distributed Systems~(TPDS), October 2016}%
%{Shell \MakeLowercase{\textit{et al.}}: Bare Demo of IEEEtran.cls for Computer Society Journals}

%\IEEEcompsoctitleabstractindextext{%

\begin{abstract}
A hybrid mobile/fixed device cloud that harnesses sensing, computing, communication, and storage capabilities of mobile and fixed devices in the field \emph{as well as} those of computing and storage servers in remote datacenters is envisioned. Mobile device clouds can be harnessed to enable innovative pervasive applications that rely on real-time, in-situ processing of sensor data collected in the field. To support concurrent mobile applications on the device cloud, a robust and secure distributed computing framework, called \maestro, is proposed. The key components of \maestro\ are (i) a task scheduling mechanism that employs controlled task replication in addition to task reallocation for robustness and (ii) \dedup\ for task deduplication among concurrent pervasive workflows. An architecture-based solution that relies on task categorization and authorized access to the categories of tasks is proposed for different levels of protection. Experimental evaluation through prototype testbed of Android- and Linux-based mobile devices as well as simulations is performed to demonstrate \maestro's capabilities. 
\end{abstract}

%\begin{IEEEkeywords}
%%Mobile device clouds; Pervasive computing; Workflows; Controlled replication; Testbed.
%%\end{IEEEkeywords}
%}

\maketitle
\thispagestyle{empty}
%\IEEEdisplaynotcompsoctitleabstractindextext
%\IEEEpeerreviewmaketitle

\section{Introduction}\label{sec:intro}
%Cite Balan
The concept of \emph{cyber foraging}---opportunistic discovery and exploitation of nearby compute and storage servers~\cite{Satyanarayanan2001}---was conceived to augment the computing capabilities of mobile handheld devices ``in the field.'' Such augmentation would enable novel \emph{compute-} and \emph{data-intensive} mobile pervasive applications spanning across multiple domains, from education to infotainment, from assisted living to ubiquitous healthcare. Such applications include (but are not limited to) object, face, pattern and speech recognition, natural language processing, and biomedical and kinematic signal processing for reality augmentation as well as for collaborative decision making. Research efforts towards realizing the vision of cyber foraging can be broadly classified into works in the fields of \emph{mobile cloud computing}, \emph{opportunistic computing}, and \emph{mobile grid computing}.
%
%Cite Serendipity, ,Viswanathan12_ICAC
Prior work in mobile cloud computing has primarily focused on augmenting the computing capabilities of mobile devices in the field with \emph{dedicated} and \emph{trusted} computing resources, either situated remotely (in the \emph{Cloud}~\cite{deng2015computation,chen2015decentralized,Chun2011,Cuervo2010,Ra2011,othman2015context}) or proximally (in \emph{cloudlets}~\cite{Satyanarayanan09,Gordon2012,AdaptiveCloudlet}). As cyber foraging was meant to convey a whole new pervasive-computing paradigm based on the principle of ``living off the land''~\cite{Satyanarayanan2001}, works in the fields of Opportunistic Computing (OC)~\cite{mtibaa2015towards,Conti2010_2,Passarella2011,Shi2012} and on mobile grid computing~\cite{Chu04,Lima2005} emerged due to the ubiquity and growing computing capabilities of mobile devices. These works have explored the feasibility of leveraging the computing and communication capabilities of other mobile devices \emph{in the field} to enable innovative mobile applications. %augment the capabilities of mobile devices and to

\begin{figure}[t!]
\centering
\includegraphics[width=3.0in]{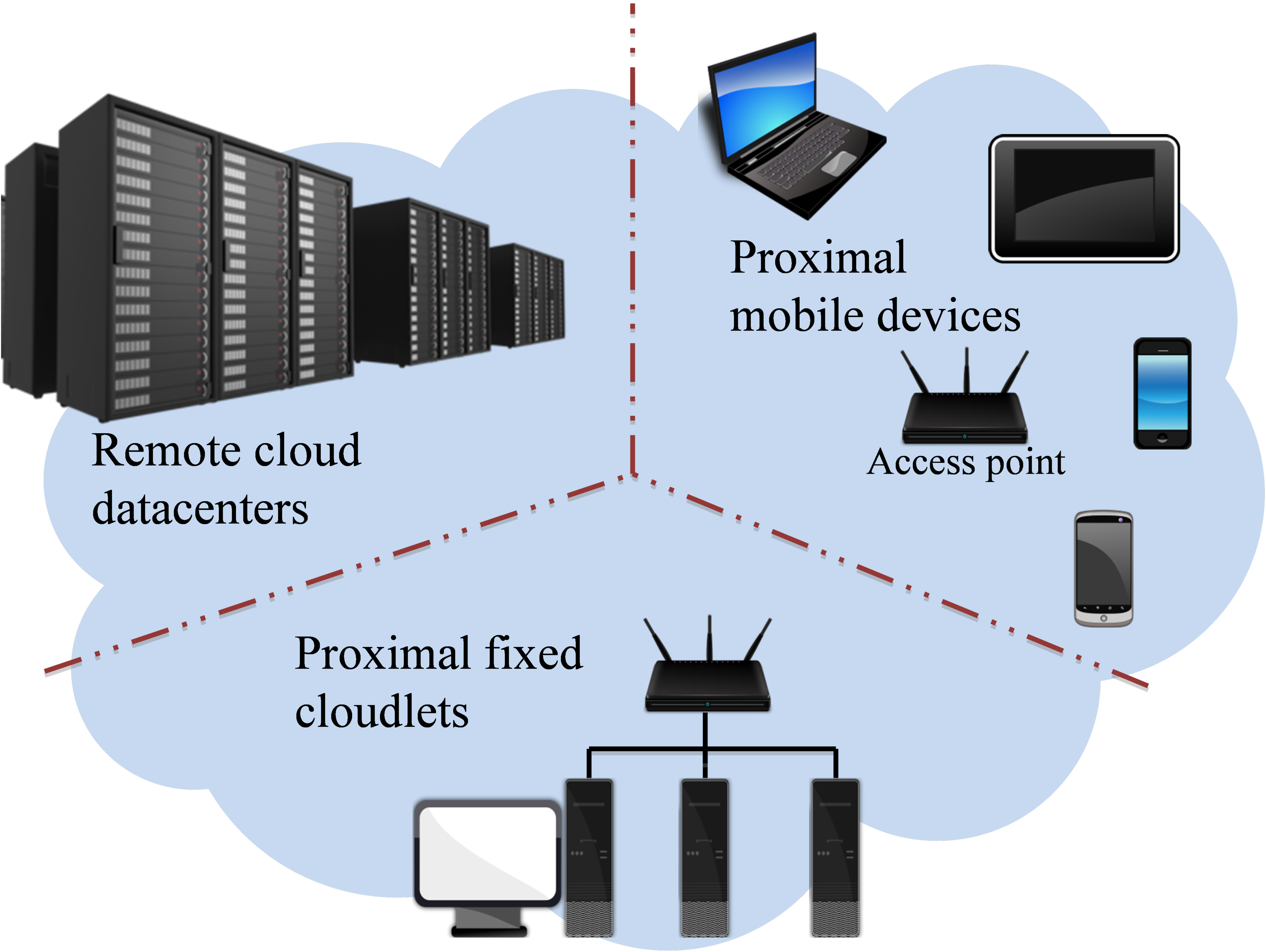}
\caption{Spectrum of computing resources in a mobile/fixed device cloud---mobile resources in the proximity, fixed (cloudlets) computing resources in the proximity usually tethered to Wi-Fi access points, and cloud resources.}\label{fig:cumulus}
\end{figure}

\textbf{Our Vision: }We envision that the heterogeneous sensing, computing, communication, and storage capabilities of mobile and fixed devices in the field \emph{as well as} those of computing and storage servers in remote datacenters can be \emph{collectively} exploited to form a ``loosely-coupled'' mobile/fixed device cloud. In keeping with the broadest principles of cyber foraging, the device cloud's computing environment may be composed of (i) purely mobile resources in the proximity, (ii) a mix of mobile and fixed resources in the proximity, or (iii) a mix of mobile and fixed resources in the proximity as well as in remote datacenters as shown in Fig.~\ref{fig:cumulus}.
%The mobile/fixed device cloud can be harnessed to enable innovative mobile applications that rely on \emph{real-time}, \emph{in-situ} processing of massive amounts of sensor data generated in the field. A role-based architecture enables easy management and does not suffer from the problem of extreme centralization. 
%In order to realize the full potential of cyber foraging.
%
%The spectrum of computing resources in a \emph{Cumulus} computing environment may range from purely mobile resources to a mix of mobile and fixed resources in the proximity to a mix of mobile and fixed resources in the proximity as well as in remote datacenters.
%which can be a data provider (a sensing device), resource provider (a computing device) or both;
We focus on the ``extreme'' scenario in which the device cloud is composed \emph{purely} of proximal mobile devices as such scenario brings all the following concerns to the fore: robustness and security. One or more of these concerns do not arise in the other scenarios and, hence, solutions developed for the extreme case will easily extend to the other cases. The device cloud employs a role-based network architecture in which the devices may at any time play one or more of the following \emph{logical roles}: (i) requester; (ii) Service (data or computing resources) Provider (SP), and (iii) broker. SPs notify the broker(s) of their capabilities and availability. The brokers are in charge of handling concurrent service requests as well as of orchestrating the execution of mobile pervasive applications on SPs. %in a robust and secure manner.

%SPs use service advertisements to notify the broker(s) of their capabilities and availability. The brokers are in charge of handling concurrent service requests as well as of orchestrating the execution of mobile applications on distributed \cumulus\ resources in a robust, secure, and privacy-preserving manner.

%\subsection{Related Work and Motivation}
%chen2003, Gordon2012
\textbf{Related Work and Motivation: }In mobile cloud computing, researchers have primarily focused on augmenting the capabilities of mobile devices in the field by offloading expensive (compute and energy-intensive) tasks to dedicated wired-grid~\cite{Hwang2004,ruth2001} or cloud resources~\cite{Chun2011,Cuervo2010,Ra2011,Gordon2012} in a transparent manner. However, these approaches are not suitable for enabling data-intensive applications in real time due to prohibitive communication cost and response time, significant energy footprint, and the curse of extreme centralization. On the contrary, we explore the possibility of mobile devices offloading workload tasks to other devices in the proximity (mobile device clouds) so to enable innovative applications that rely on real-time, \emph{in-situ} processing of sensor data.
%
%Conti2010_1, ,Viswanathan12_ICAC
Research in the areas of mobile device clouds~\cite{Chu04,Lima2005} and OC~\cite{Conti2010_2,Passarella2011,Shi2012} has explored the potential of code offloading to proximal devices by following two different approaches. Solutions for mobile grids advocate a structured and robust approach to workflow and resource management; whereas OC depends entirely on direct encounters and is highly unstructured with little or no performance guarantees. These works do not address the crucial research challenge of ``real-time'' concurrent applications management while also taking the following concerns into account: robustness, security, and privacy.
%In this paper, we built on the former approach to address the crucial research challenge of concurrent workflows management in real-world mobile device clouds.

Management of concurrent workflows has been studied before in the context of wired-grid computing. In~\cite{Bittencourt2010,Stavrinides2011,Tsai2012}, different strategies for scheduling multiple workflows were investigated---namely, sequential, by interleaving tasks of the different workflows in a round-robin manner, and by merging the different workflows into one. Independently of the strategy, the workflow tasks are allocated using level-based~\cite{Stavrinides2011}, list-based~\cite{Bittencourt2010}, duplication-based~\cite{Park1997}, or clustering-based~\cite{Tsai2012} heuristics. In all these heuristics, all the tasks of a workflow are scheduled at the same time with the option of filling the ``gaps" (schedule holes due to high communication cost between tasks) for efficient resource utilization.

However, none of the existing solutions can be adopted for workflow management mobile device clouds as they do not factor in task deduplication (for efficiency), reactive self-healing and proactive self-protection (for failure handling), security (from malicious resources), and privacy, which are \emph{all} primary concerns in mobile cloud computing. Even though duplication-based scheduling provides some level of redundancy, it treats only fork tasks (ones with multiple successor tasks) as critical and does not protect other tasks that may be critical in the context of the application (or annotated by the developer or user as one).  All the aforementioned shortcomings serve as a motivation for the clean-slate design of \maestro. To the best of our knowledge, ours is the first work to explore deduplication and scheduling of tasks belonging to concurrent real-time application workflows on mobile device clouds.

%the problem of concurrent workflow management in mobile device clouds.
%for concurrent workflows management in \cumulus\ (mobile device clouds) with the desired properties.

%Prior work on workflow scheduling in grid environments. Their assumptions and what they lack? \newline
%Prior work on meta brokers in grid environments. Their assumptions and how they are similar/different to our solution for inter-arbitrator coordination? \newline

%Cite \cite{Viswanathan12_ICAC} for role-based middleware. Mention the properties it has and what it is missing. \newline
%Cite \cite{Viswanathan12_ETSIoT} for self-healing property and how it is a fundamentally different approach to self-protection. \newline
%Criticality necessitates proactive self-protection. Healing is reactive and may delay service response. \newline
%Idea: Controlled redundancy. XX Splitting tasks into micro-tasks for homogeneity in task sizes leads to a lot of dependencies. Failures cannot be tolerated. Healing is slow. Hence, redundancy. \newline
%1) Criticality, 2) More complex workflow, i.e., more dependencies, 3) More tasks to complete and more data to process.
%Cite [XX,XX] papers on opportunistic computing to talk about prior work in self-protection. Highlight the novelty, controlled redundancy, and at a very high-level description of what the philosophy is. \newline

%\subsection{Contributions}
\textbf{Contributions: }We present \maestro, a framework for robust and secure mobile cloud computing where any mobile application is represented as a workflow, i.e., a Directed Acyclic Graph (DAG) composed of multiple parallelizable and sequential tasks along with dependencies. Often, multiple service requests are received \emph{simultaneously} by brokers and, hence, tasks belonging to multiple workflows managed by the brokers have to be allocated and executed on the SPs in the mobile device cloud. %We would like to emphasize that the concurrent workflows presented in this paper represent {\it futuristic} applications that open up a variety of challenges exclusive to concurrent workflows. In this paper, we have explained these challenges and presented novel solutions for them.
%
%For example, effective hypoxia (i.e., lack of oxygen) detection in people at high altitudes requires simultaneous assessment of multiple factors like context, physical or psychological stress as well as environmental conditions. Assessment of each of the aforementioned factors relies on in-situ real-time processing of heterogeneous sensor data (ranging from environmental to kinematic to vital-sign data).
%
\maestro\ employs controlled \emph{replication of critical workflow tasks} and controlled \emph{access to user data} (via multiple levels of authorization) to realize \emph{secure} mobile computing. Controlled replication refers to the idea of replicating ``critical'' workflow tasks and only when needed (based on SPs' reliability). Not only task replication imparts \emph{robustness} (against SP failures); it has the ability to handle \emph{uncertainty} arising from device failures, denial of service, and intentional corruption of results. We categorize the workflow tasks according to the sensitivity of the data processed, and allow access only to authorized service providers.
% but also provides \emph{security} (from malicious nodes masquerading as SPs). Therefore, \maestro's task scheduling is \emph{Byzantine fault tolerant}~\cite{Castro1999}, i.e.,

Note that, while concurrent workflows makes the real-time task-allocation problem complex, it also presents opportunities: there may be multiple duplicate service requests at a broker; similarly, there may be multiple duplicate tasks that are common across workflows. \maestro\ \emph{deduplicates similar tasks across workflows} as it lends itself to minimization of duplication in services rendered. Task deduplication leads to efficient real-time, in-situ processing of simplified workflows (with fewer tasks than before) as well as to better utilization of resources. To address the non-trivial research challenge of the identification of task duplicates and the creation of simplified workflows (at the brokers), we introduce \dedup, a sub-graph matching technique for task deduplication among DAGs. After deduplication, the tasks of the simplified workflows have to be scheduled for execution on the mobile/fixed cloud resources so to meet the user-specified deadline of the corresponding application workflow.

To sum up, \emph{our contributions} are: 
\begin{itemize}
\item We present a robust and secure mobile cloud computing framework, \maestro, for concurrent application workflow management on mobile device clouds; 
\item We impart robustness to \maestro\ via two core components, namely, 1) replication-based task-scheduling mechanism and 2) \dedup, for task deduplication among concurrent DAGs. 
\item We discuss the results from our experimental evaluation of \maestro\ in representative operating environments. We also present analysis of our replication-based task-scheduling mechanism for mobile device clouds through experiments on a prototype testbed of Android- and Linux-based devices.
\end{itemize}
%Simulations were performed on a purpose-built Java$^{\textsc{TM}}$-based simulator. 

\begin{figure*}[t!]
\centering
\begin{tabular}{ccc}
\includegraphics[width=1.35in]{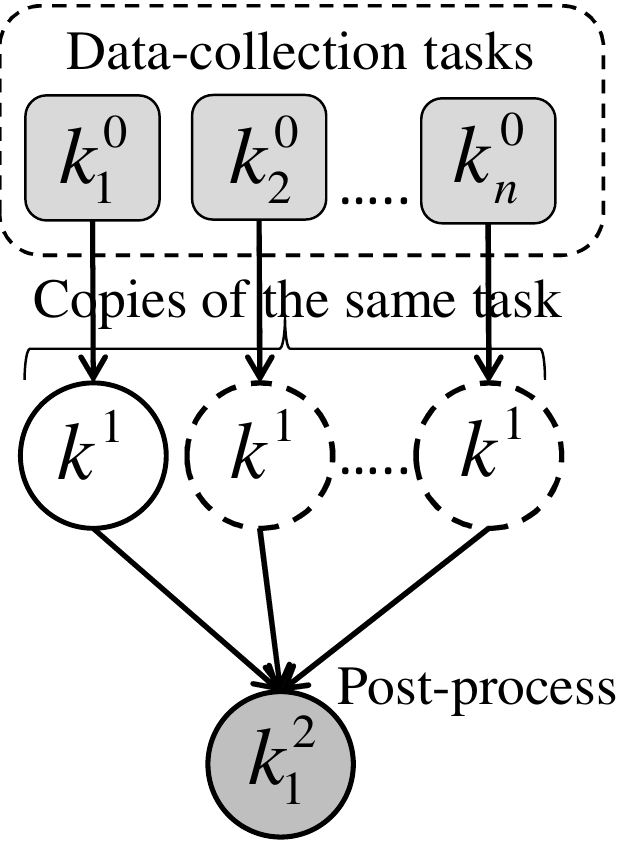} &
\hspace{1cm}
\includegraphics[width=1.3in]{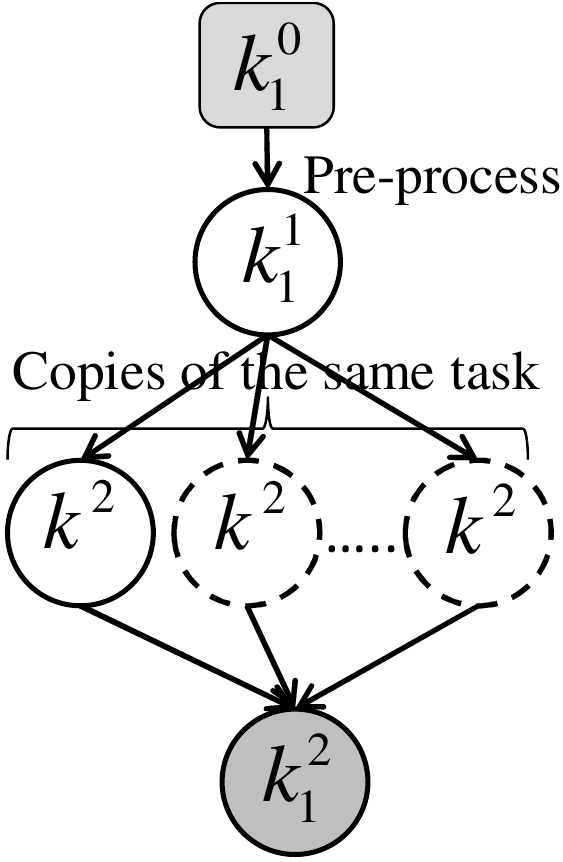} &
\hspace{1cm}
\includegraphics[width=2.4in]{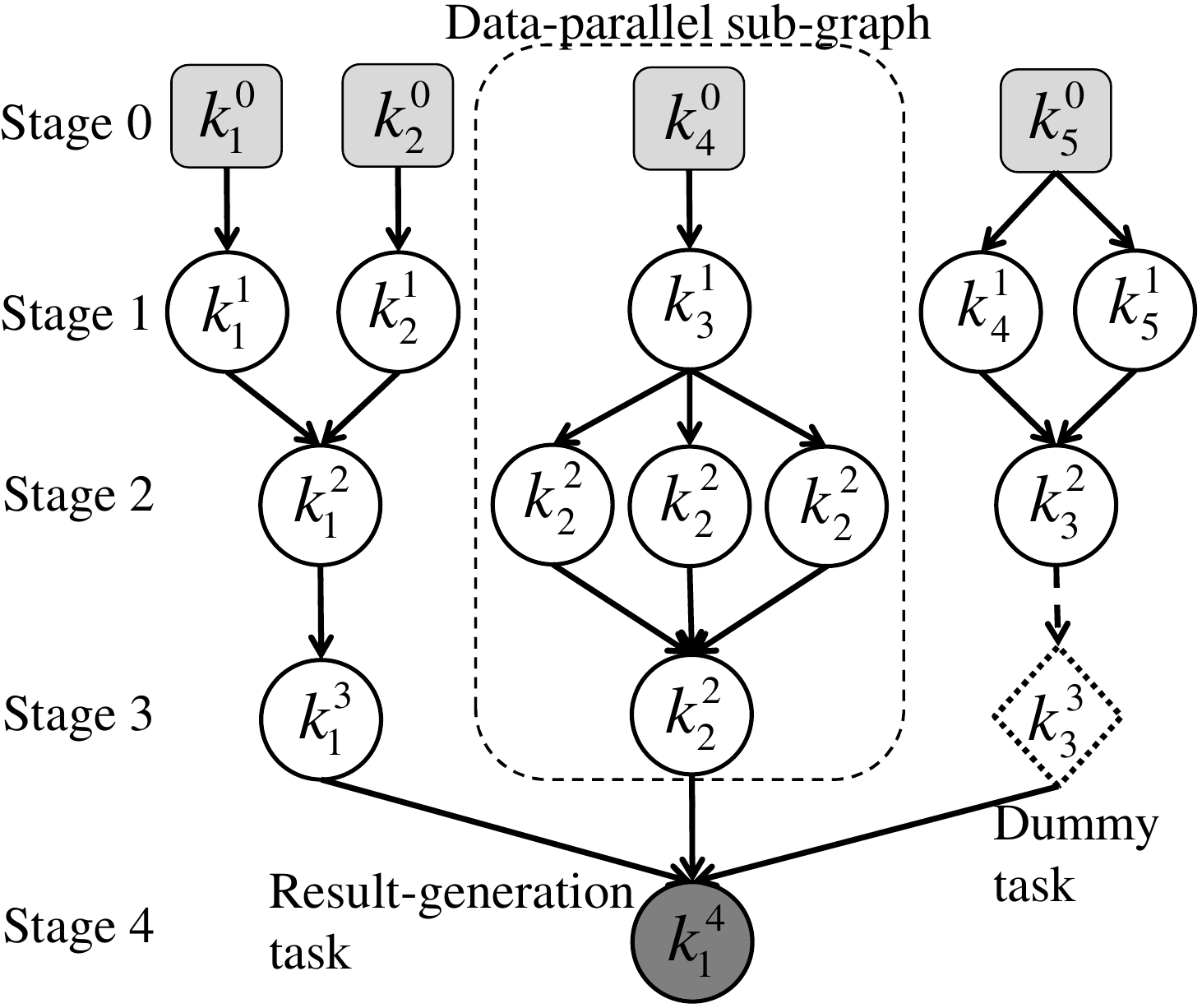} \\
\small(a) Data-parallel & \small(b) Data-parallel & \small(c) Task-parallel
\end{tabular}
\caption{Workflows for \emph{data-parallel} applications with $n$ parallel tasks (a) when there are $n$ separate data sources and (b) when data from a single source has to be divided into multiple chunks (pre-processing); (c) Workflow for a \emph{task-parallel} application (with a data-parallel sub-graph in it)---here a ``dummy task" is also represented in Stage~3 of the right branch.}\label{fig:workflows}
\end{figure*}

\textbf{Outline:} In Sect.~\ref{sec:cumulus}, we present the architecture of mobile device clouds and our generalized workflow representation scheme; in Sect.~\ref{sec:maestro}, we discuss the task scheduling mechanism for concurrent workflows and task deduplication algorithm to handle duplicate tasks in concurrent workflows; in Sect.~\ref{sec:evaluation}, we discuss quantitative results that demonstrate the merits of our contributions; finally, in Sect.~\ref{sec:conclusions}, we conclude with a note on future work.
\section{Mobile Device Cloud and Workflows}\label{sec:cumulus}
%In this section, we introduce the necessary background pertaining to mobile application workflows and mobile device clouds so i) to recap the terminology introduced in our prior work on mobile cloud computing, ii) to aid in the understanding of our contributions, and iii) to differentiate the contributions of this paper from our prior work.

%The tasks are elementary and cannot be split further into micro-tasks. Each task is either in itself or a building block of a computational model that aids in the extraction of information from data.
In this section, firstly, we introduce a mobile device cloud, our envisioned heterogeneous computing environment; then, we discuss a generalized workflow representation scheme for depicting the data-processing chains in mobile applications.

%\subsection{Cumulus}
%\todo{an introductory sentence before starting}
%As mentioned earlier,
\subsection{Mobile Device Cloud}
Our vision is to organize the heterogeneous sensing, computing, and communication capabilities of mobile devices in the proximity (as well as in remote datacenters) in order to form an \emph{elastic resource pool}---a mobile device cloud (MDC). This cloud can then be leveraged to augment the capabilities of any mobile device in the network when required in order to enable novel mobile applications.  The interested reader may refer to our work in~\cite{Viswanathan12_TPDS} for details on the architecture of MDCs.

\emph{\underline{Logical roles}: }MDC is a hierarchical logical-role-based computing environment in which the devices may play one or more of the following logical roles: (i) \emph{requester}, which place requests for application workloads that require additional data and/or computing resources from other devices, (ii) \emph{service provider}, which can be a data provider (a sensing device), a resource provider (a computing device) or both; (iii) \emph{broker}, which is in charge of handling requests and orchestrating the execution of applications on the MDC. This architecture enables easy management and does not suffer from the problem of extreme centralization (i.e., single point of failure) as these are only logical roles.

\emph{\underline{Service providers and broker}: }The spectrum of computing resources (SPs) in a MDC computing environment includes mobile (sensing and computing) resources in the proximity, fixed (dedicated cloudlets) computing resources in the proximity usually tethered to Wi-Fi access points or base stations, and fixed resources in remote datacenters (dedicated cloud resources). %as shown in Fig.~\ref{fig:cumulus}.
As mentioned earlier, we  design solutions capable of handling the extreme case, i.e., a hybrid cloud composed \emph{only} of proximal mobile computing resources. Another design decision that is faced with a spectrum of possibilities is the location of the broker. The role of a broker can be played by one of the mobile resources\footnote{Broker selection %in a purely ad hoc network of mobile devices
is out of the scope of this article.} (chosen based on centrality of location, battery level, and/or computing capabilities) or be one of the proximal fixed resources.
%\rev{address the multiple broker problem here??}

\emph{\underline{Service discovery}: }The role of broker is played by one of the proximal fixed resource tethered to the Wi-Fi access point or the base station so to ensure that all SPs are connected to a broker when they are in the network. Service discovery at the broker is achieved through service advertisements from the SPs. Service advertisements may include information about the sensor data, types of sensors (quality of data), amount of computing (in terms of unutilized CPU cycles~[\%]), memory ([$\mathrm{Bytes}$]), and communication ([$\mathrm{bps}$]) resources, the start and end times of the availability of those resources, and the available battery capacity ([$\mathrm{Wh}$]) at each SP. The broker leverages this information to allocate workload to the SPs. %\todo{first, we say broker. Then, we say at the brokers...is it one or multiple? if one, we do have single point of failure; if it's multiple, how do we handle them? A sentence is needed to clarify that it's the latter but somehow the handing problem is outside the focus of this paper, and add a reference.}
%\todo{Talking about multiple brokers here opens a can of worms. It just becomes another useless negative comment where the reviewer will claim we have not elaborated on a ``crucial'' aspect of the mobile cloud computing problem. It is however smart to put it in the future work section. That way, we do not claim that this problem is simply out of ``scope'' but a bigger problem that needs a more elaborate treatment.}

\emph{\underline{Uncertainty awareness}: }The broker extracts the following long-term statistics from its underlying resource pool: the average arrival (joining) rate of SPs ($\widetilde{W}$), the average SP availability duration ($\widetilde{T}$), and the average number of SPs associated with the broker at any point in time ($\widetilde{N}$) whose relationship is given by Little's law%~\cite{Little61}
, i.e., $\widetilde{N}=\widetilde{W} \cdot \widetilde{T}$. These statistics help the brokers assess the \textit{churn rate} of SPs, i.e., a measure of the number of SPs moving in to and out of their respective resource pools over a specific period of time. When the long-term statistics are not taken into account at the broker and when the durations advertised by the SPs are used to make workload allocation decisions, the mismatch between advertisements and ground reality will have an adverse effect on the performance of applications (particularly, in terms of response time). Uncertainty awareness at the broker enables design of robust workload scheduling algorithms. %Note that the brokers need not be aware of the underlying probability distribution of SP arrivals.

%Churn rate of service providers will be different in different geographic location. For example, the churn rate of service providers at a shopping mall is far greater than the one at a coffee shop. Also, at a particular location (say, the coffee shop), the churn rate can vary over time (e.g., depending on the time of the day).

%When the churn rate of service providers is high, i.e., the average duration of service providers availability is low, the percentage of potentially costly migrated workload tasks will be high if the resource-allocation engine does not possess uncertainty awareness.
%The broker is aware of the instantaneous power drawn by the workload tasks of a specific application when running on a specific class of CPU and memory (together given by $c^{comp}_n~[\mathrm{W}$]) as well as network ($c^{net}_n~[\mathrm{W}$]) resources at each service provider as the information about the different types of devices is known in advance.

%The self-optimization solutions we have developed so far for task allocation in data- and task-parallel (XX PROVIDE DEFINITION XX) mobile applications on MDC focused on management of a ``single'' workflow with the aim of minimizing the makespan (total execution time) and/or the maximum battery drain among MDC resources.

%\subsection{Generalized Workflow Representation}
%\todo{an introductory sentence before starting}
\subsection{Workflow Representation}
Applications are composed of tasks whose order of execution is specified by workflows.

\emph{\underline{Structured workflows}: }Our generalized workflow is a \emph{Directed Acyclic Graph (DAG)} composed of tasks (vertices) and dependencies (directed edges), as shown in Fig.~\ref{fig:workflows}. Tasks belong to one of the following three categories: 1) data-collection task, 2) computation task, or 3) result-generation task. These tasks are elementary and cannot be split further into \emph{micro-tasks}, i.e., parallelization of elementary tasks does not yield any speed-up in execution time. The workflow is composed of multiple \emph{stages} with a set of tasks to be performed at each stage. The resources that perform the tasks at Stage~$i-1$, where $i\geq1$, serve as data sources for the tasks that have to be performed at Stage~$i$. The data sources for tasks at Stage~$1$ are the sensors themselves (where Stage-$0$ tasks, i.e., data-collection tasks, are performed).
There are no dependencies between the tasks at a particular stage and, hence, they can be performed in parallel. Also, without any loss in generality, we assume that there can be dependencies only between tasks of consecutive stages in the workflow. Whenever we have dependencies between tasks of non-consecutive stages, we introduce the notion of ``dummy tasks,'' whose output equals the input and whose cost of operation (in terms of time and battery drain) is zero. Our \emph{structured} workflow representation is rich in information. In addition to the regular information like task types and data dependencies, it also includes the following: task identifiers, task sizes, quality and quantity of inputs, the preferred interfaces to child and parent tasks, the implementation (when multiple exist), and the task criticality (either boolean or multiple degrees).
%Figure~\ref{fig:hypoxia_workflow} shows a 5-stage workflow for hypoxia detection (lack of oxygen) represented using our generalized workflow representation.

\emph{\underline{Concurrent workflows}: }Concurrent service requests are often received by each broker, i.e., multiple workflows have to be executed concurrently in the underlying pool. The aforementioned task-allocation problem, albeit complex, presents opportunities. There may be multiple similar service requests at a broker as well as multiple tasks that are common across different workflows. Deduplication of such common tasks leads to efficient real-time, \emph{in-situ} processing of simplified workflows (with fewer tasks than before) as well as to better resource utilization.%\todo{DARIO: again, are we envisioning multiple brokers receiving the same service request or only one? we can say that former is that general case and that after some `request consolidation' across multiple active brokers, we get to Fig.~\ref{fig:applications}; anyway, inter-broker comms and coordination are needed to avoid single point of failure}.

%\todo{I would again prefer discussing multiple broker scenario in the future work section (which we can call ``discussion'' section. It would be a very nice touch to the paper. Rather than raising these questions ourselves in the middle without complete answers. Parul, please take text from my last TPDS journal for discussion on multiple broker scenarios. Also, reuse our early text on how we can deduplicate across brokers. I remember us writing something.)}

\begin{figure*}[t!]
\vspace{-1.30in}
\centering
\begin{tabular}{ccc}
\includegraphics[width=2.3in]{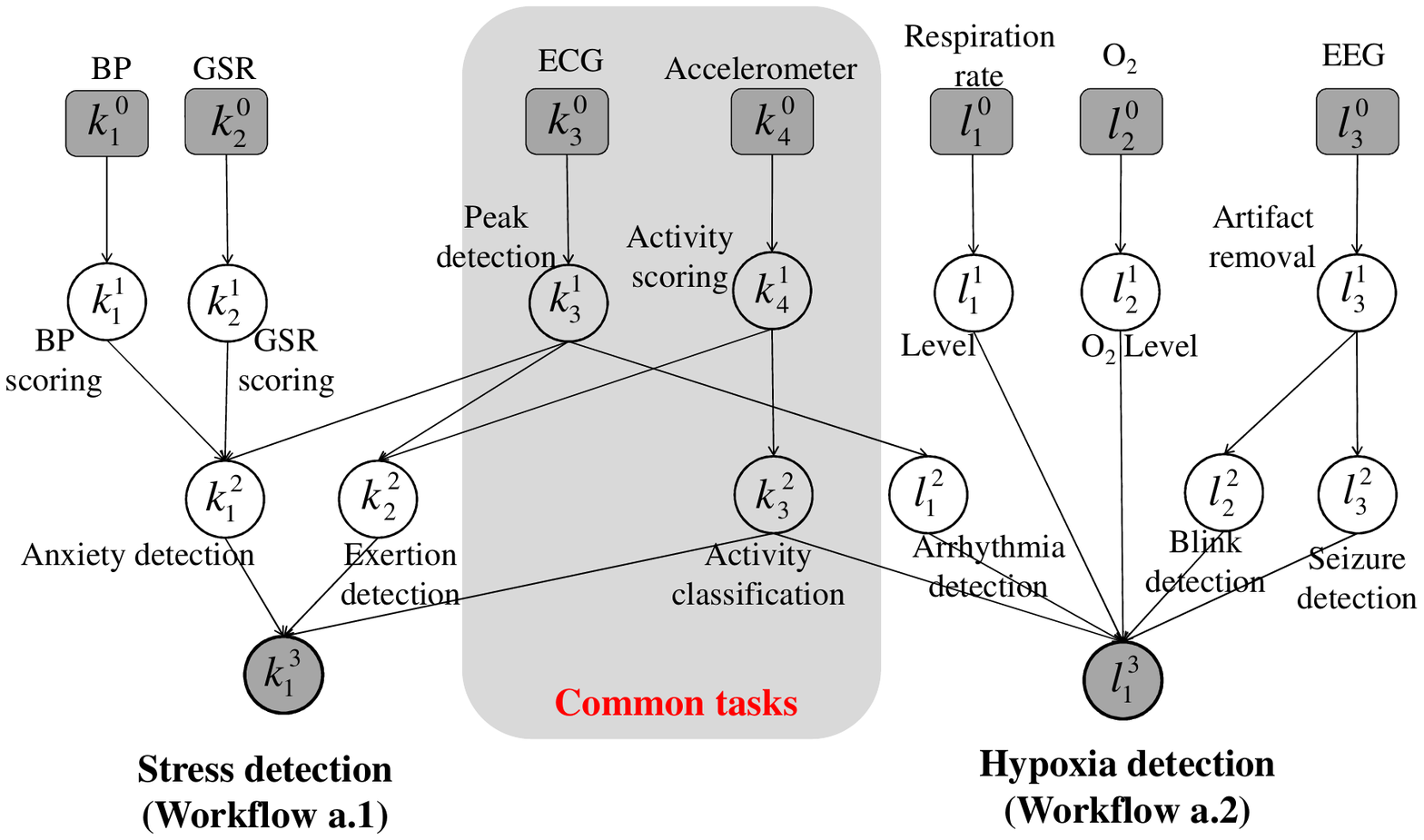} &
\includegraphics[width=2.20in]{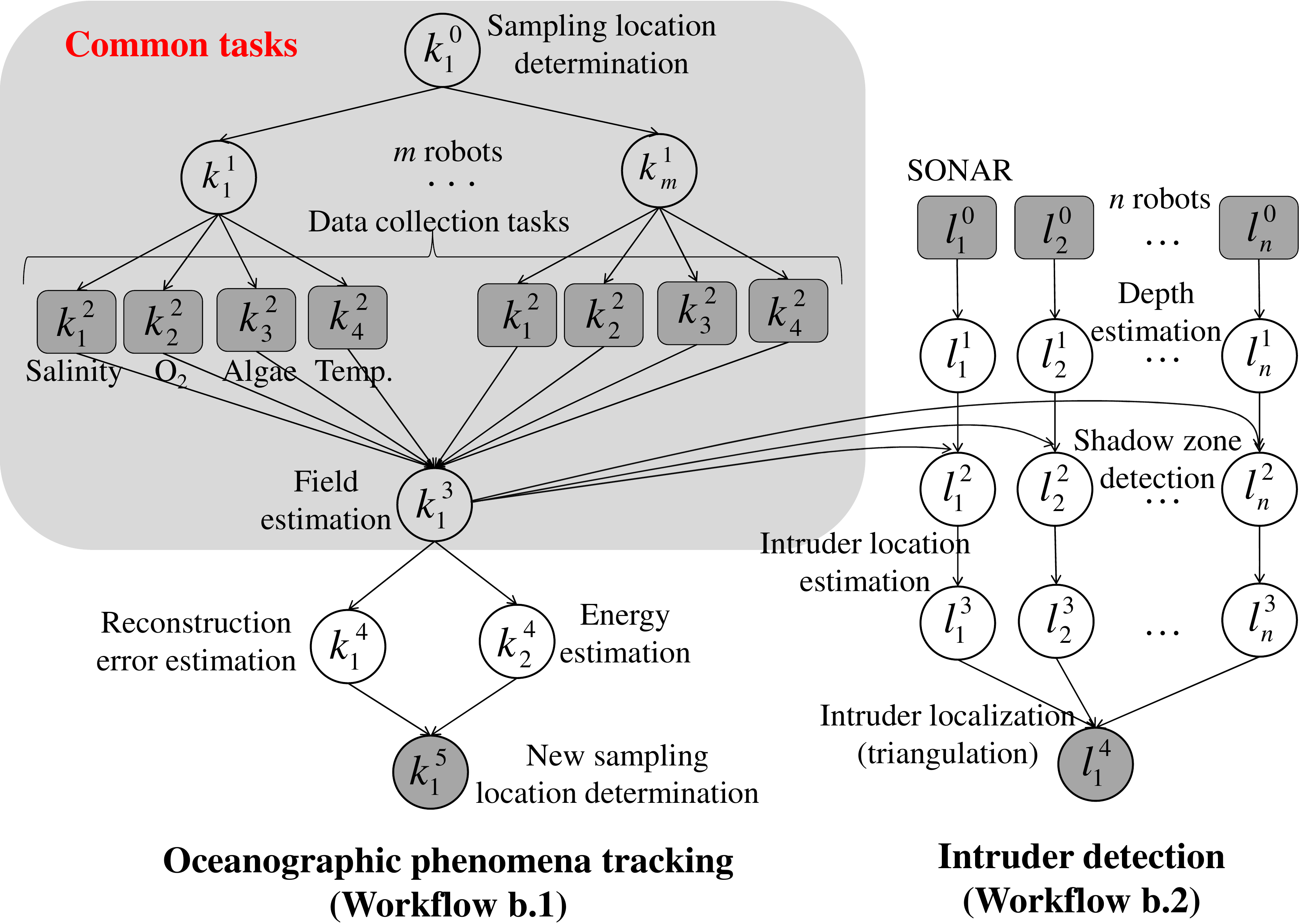}&
\includegraphics[width=1.50in]{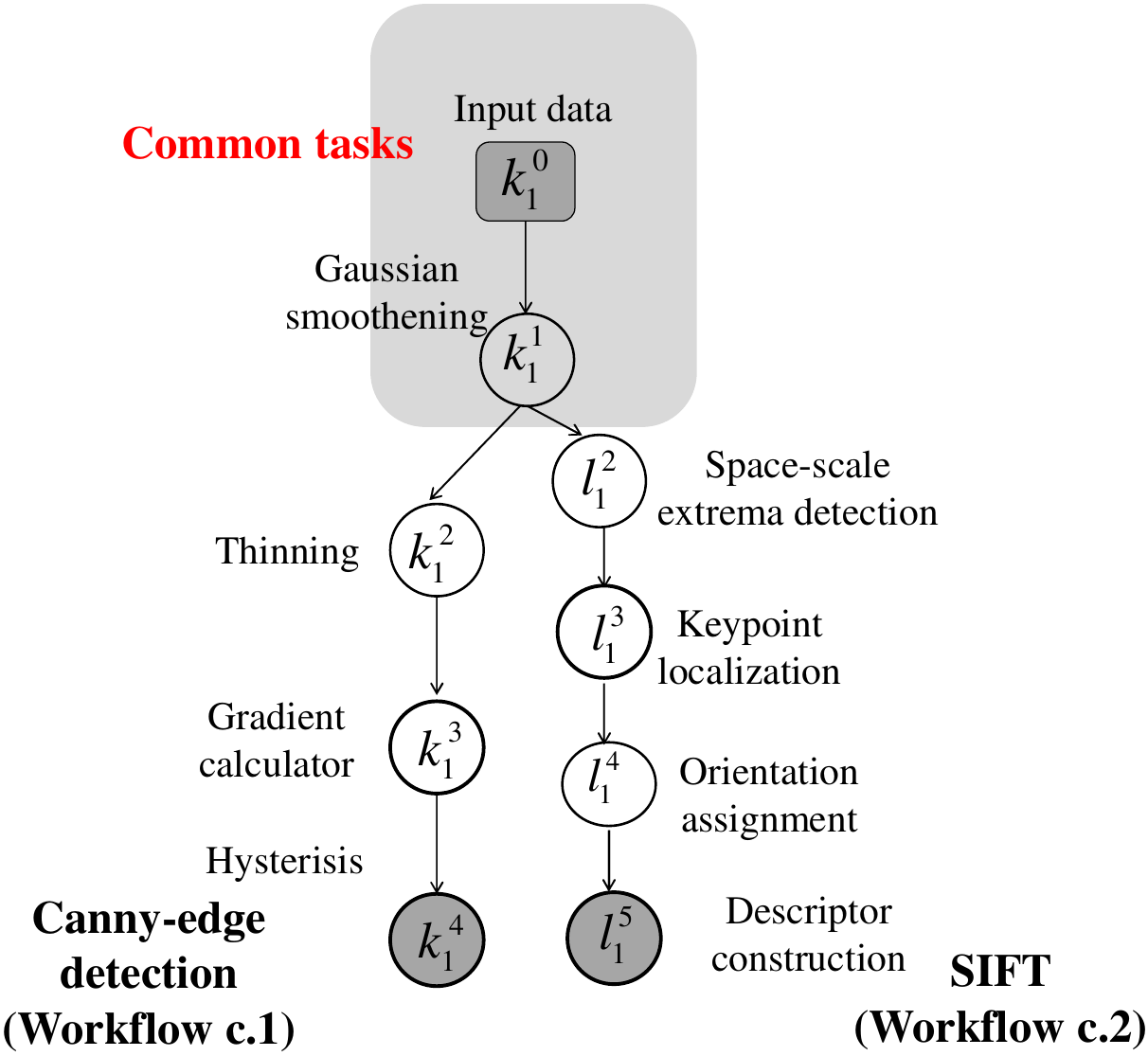} \\
\small (a) Ubiquitous healthcare domain & \small (b) Distributed robotics domain & \small (c) Computer vision domain
\end{tabular}
\caption{Example workflows from two different domains, ubiquitous healthcare (a), distributed robotics (b), and computer vision domain (c). Common tasks across workflows in the same domain are highlighted. Workflows in (a) are purely task parallel, in (c) purely data parallel, while the ones in (b) are mixed. }\label{fig:applications}
\end{figure*}

\emph{\underline{Example applications}: }This representation is powerful as it captures both \emph{data-} and \emph{task-parallel} applications. Data-parallel applications are also referred to as  ``embarrassingly parallel'' applications, in which an independent set of homogeneous tasks---working on disjoint sets of data---can be performed in parallel (preceded and succeeded by pre- and post-processing tasks, respectively), as shown in Figs.~\ref{fig:workflows}(a) and \ref{fig:workflows}(b). Task-parallel applications, on the other hand, have a set of sequential as well as parallel tasks with pre-determined dependencies and degree of parallelism. A task-parallel workflow may also have a data-parallel block built into it, as illustrated in Fig.~\ref{fig:workflows}(c). To understand the concept of workflows and their constituent computational tasks, consider the following example applications from different domains.

\emph{\textbf{(a)}~Ubiquitous healthcare. }Applications in this domain (also called data-driven, sensor-based healthcare) include stress detection, hypoxia (lack of oxygen) detection, alertness and cognitive performance assessment. Figure~\ref{fig:applications}(a) depicts the task-parallel workflows for \emph{stress detection} (Workflow~a.1) and \emph{hypoxia detection} (Workflow~a.2), which use vital-sign data acquired from biomedical (e.g., Blood Pressure (BP), ElectroCardioGram (ECG), ElectroEncephaloGram (EEG)) as well as kinematic sensors (e.g., accelerometer, gyroscope) attached to a person. Dummy tasks are not shown for the sake of simplicity. The tasks in these two workflows belong to one of the following classes: data analysis, data manipulation, decision making; and they aid in determining the psychophysiological state of a person (knowledge) from raw sensor data. 
As mentioned earlier, applications belonging to the same domain may have similar component tasks, which can be deduplicated to achieve efficiency. For example, feature extraction from accelerometer outputs as well as ECG analysis (in Stage~1) are component tasks in the hypoxia-detection workflow (for context assessment, e.g., activity and arrhythmia detection, in Stage~2) as well as in the stress-detection workflow (for exertion detection in Stage~2), as shown in Fig.~\ref{fig:applications}(a).

\emph{\textbf{(b)}~Distributed robotics. }Distributed decision-making applications in this domain include adaptive sampling, intruder detection, target tracking, data-driven path/trajectory planning, to name just a few. Figure~\ref{fig:applications}(b) depicts workflows for \emph{oceanographic phenomena tracking} (Workflow~b.1) and coastal underwater \emph{intruder detection} (Workflow~b.2), which use all or a subset of the following data acquired using environmental and inertial navigation sensors as well as SONAR on autonomous underwater robots: temperature, salinity, pollutants, nutrients, position, and depth. Workflow~b.1 depicts how field estimation is used to track the effect of oceanographic phenomena (e.g., temperature and salinity gradients, algae growth, nutrient concentration) on aquatic life. Workflow~b.2 depicts intruder localization (using SONAR), which requires optimal positioning of the robots in order to avoid false positives due to severe transmission losses of the acoustic signal/waves traversing certain regions. Such regions of high transmission loss can again be determined from temperature, salinity, and depth field estimates. 
Here, field estimation is a common task (between the two workflows), which can be deduplicated. Note that in this application the overall task-parallel workflow is composed of smaller data-parallel workflows.

\emph{\textbf{(c)}~Computer vision. }Many applications in this domain such as object recognition, face recognition, gesture recognition, and image stitching use two well-known algorithms shown in Fig.~\ref{fig:applications}(c), \emph{Canny edge detection} (Workflow~c.1) and \emph{Scale Invariant Feature Transform (SIFT)} (Workflow~c.2). Workflow~c.1 takes as input an image and uses a multi-stage algorithm to detect edges in the image. Workflow~c.2 generates a large collection of feature vectors from an image, each of which is (i) invariant to image translation, scaling, and rotation, (ii) partially invariant to illumination changes, and (iii) robust to local geometric distortion. Both these workflows have a common task, Gaussian smoothening, which can be deduplicated. 

%Again, the tasks in these two workflows belong to one of the aforementioned three classes.

\section{Maestro}\label{sec:maestro}
In this section, we present \maestro, a robust mobile cloud computing framework for concurrent workflow management on the MDC. Firstly, we present a concurrent workflow-scheduling mechanism designed for \maestro. Secondly, we discuss \maestro's task-scheduling mechanism, which employs controlled task replication (for robustness) before scheduling the tasks for execution on the MDC's resources. At the end, we present \dedup\ (the sub-graph matching technique in \maestro) for task deduplication among DAGs.

\subsection{Concurrent Workflows Scheduling}
%\textbf{Concurrent Workflows Scheduling: }%After deduplication, the simplified workflows arrive at the broker.
The brokers receive multiple workflow execution requests over a period of time from the service requesters. The tasks of these workflows have to be allocated to SPs in the MDC. While submitting a service request, the application can specify the \emph{absolute deadline} ($D~\mathrm{[s]}$) within which the workflow execution has to be completed for it to be useful. There is also a notion of an acceptable probability of failure ($P^{fail}$) for each workflow. This probability can be a service-level guarantee advertised by the broker or negotiated a priori between brokers and service requesters. \maestro's task-scheduling mechanism at the broker is in charge of determining (i) the set of workflow tasks that are ready to be allocated, (ii) the relative priority among the ready tasks\footnote{A ready task is one that does not have any unresolved dependencies, i.e., all its parent tasks have completed their execution.}, and (iii) the amount of replication and the appropriate SP(s) for each ready task.
In \maestro, tasks can be immediately allocated \emph{as and when} they become ready \emph{or} the ready tasks can be accumulated (over a waiting period, $\delta^{ready}_b~\mathrm{[s]}$) and then allocated for a more efficient schedule in terms of \emph{makespan} (i.e., total workflow execution time) and number of replicas (i.e., battery drain). This waiting period is again a tunable parameter; the larger the waiting period, the greater the chances of finding the most appropriate SPs (lower makespan and fewer replicas). However, $\delta^{ready}_b$ cannot be too large due to real-time constraints of the application.

\textbf{Task Prioritization: }Determining the relative priority among ready tasks from the same or different workflows requires incorporation of computation-time information and deadline requirements, as discussed in prior work on workflows management in computational grids~\cite{Stavrinides2011}. Firstly, we determine the \emph{level} of task, which is the length of the longest path from that task to an exit task. The length of a path in a DAG is the sum of the average computation time of that task and the average computation times of all the successor tasks along the path. The average communication times between successive tasks should also be taken into account if the Communication-to-Computation-costs Ratio~(CCR) of the workflow DAG is high. The level $\Delta^k~\mathrm{[s]}$ of a task $k$ is given by, 
\begin{equation}
\Delta^k = \overline{\alpha^k} + \max_{c \in \mathcal{C}^k} \{\overline{\beta^{kc}} + \Delta^c\},
\end{equation}
where $\overline{\alpha^k}$ is the average computation time of a task $k$ on the SPs in the MDC, $\mathcal{C}^k$ is the set of child tasks of $k$, and $\overline{\beta^{kc}}$ is the average communication time for data transfer between tasks $k$ and $c$ when executed on the SPs in the MDC. Once the level of each ready task is known, their \emph{slack} $S~\mathrm{[s]}$ (maximum allowable wait time before execution of that task) at any time $t$ is determined, for task $k$, as, 
\begin{equation}
S^k(t) = D^k - \Delta^k - t,
\end{equation}
where $D^k$ is the absolute deadline for the workflow which task $k$ belongs to. The ready task $k^*$ with the \emph{smallest} slack has the \emph{highest} priority, i.e., $k^*=\arg\min_k{S^k(t)}$.

After prioritization of the ready tasks according to this criterion, the most appropriate SP for allocation and the amount of replication (when necessary) are determined. Each SP in the MDC has a task queue. For a ready task $k$ with the highest priority, the SP $n$ that provides the earliest finish time ($t^{k,fin}_n$) is the most preferred. Finish times are considered due to heterogeneity in capabilities of service providers. In a homogeneous environment, start times are sufficient to make allocation decisions. The $t^{fin}$'s are obtained as,
\begin{equation}
t^{k,fin}_n = t^{k,start}_n + \alpha^k_n,
\end{equation}where $t^{k,start}$ is the start time for task $k$ on SP $n$. The $t^{start}$ depends on the number and type of existing tasks in the queue. However, there is uncertainty associated with the availability of the SPs in a MDC for the required duration and this has to be taken into account in the scheduling mechanism.

%\subsection{Controlled Replication}
\textbf{Controlled Replication: }An effective way to overcome the uncertainty (due to failures) is the reallocation of failed tasks (also called ``healing''). However, healing is \emph{not} suited for tasks with large computation times and for tasks that are critical for multiple workflows. Though healing provides robustness, it does increase the makespan as it waits for at least the task's computation time before making a decision (i.e., it is \emph{reactive}). Conversely, we replicate \emph{critical} tasks at multiple service providers (\emph{proactively}) in order to ensure the completion of those tasks on time. Proactive task replication avoids unnecessary idle waiting times incurred in reactive failure handling (i.e., healing). Tasks that have to be replicated are allocated to the SP that provides the next earliest $t^{k,fin}$. Note that, as replicas have the same priority as the original, they are allocated together with the original before the other tasks that have lower priority.

All tasks in a workflow should not be replicated as it will increase the total number of tasks to be executed, in turn leading to massive queuing delays and large makespans. The application developer may explicitly \emph{annotate} certain workflow tasks as \emph{non critical}. All other tasks are treated as \emph{blocking tasks}, i.e., the progress of the workflow depends on their completion. The decision to replicate a task $k$ initially allocated to SP $n$ is taken based on how the task-completion probability of $n$ compares with the ``required'' success probability for that task derived from the pre-specified $P^{fail}$. The required success probability $p^{succ}$ for each task in the set of incomplete tasks $\mathcal{K}$ of a workflow is obtained by solving,
\begin{equation}
(p^{succ})^{|\mathcal{K}|} = 1-P^{fail}.
\end{equation}

The task-completion probability $p^{k,succ}_n$ of a task $k$ at SP $n$ is, 
\begin{equation}
p^{k,succ}_n = \Pr\{t+T^n > t^{k,start}_n+\alpha^k_n\},\label{eq:protec}
\end{equation}where $t$ is the current time and $T^n$ is the ``actual'' availability duration of SP $n$. Without any loss in generality, we assume that the distribution of SP availability duration is known while determining $p^{k,succ}_n$. It is quite straightforward to obtain and maintain such statistics at the brokers. When $p^{k,succ}_n<p^{succ}$, a replica is allocated to the next best SP as mentioned before. Replicas of task $k$ are created and allocated to SPs until the following condition is satisfied for the first time , i.e.,
\begin{equation}
1 - \prod_{n \in \mathcal{N}} (1 - p^{k,succ}_n) \geq p^{succ},
\end{equation}where $\mathcal{N}$ is the set of SPs which the replicas are allocated to. As the tasks of a workflow are completed over time with probability one, the required success probability of remaining incomplete tasks decreases; this allows the scheduler to use some less reliable SPs, resulting in load balancing.

For ``fork'' tasks that are common across multiple workflows, the more stringent condition on the required probability of success is taken into account. To avoid uncontrolled replication, we use a maximum replication limit. This limit is different for different types of tasks. For example, the fork tasks, which are crucial for the success of multiple workflows, have more replicas than the other tasks. This difference in the level of protection is crucial to avoid blocking of MDC resource by tasks that are not so critical as the ones that follow them.

\emph{\underline{Levels of protection}: }
%
%\begin{figure}[t!]
%\centering
%\includegraphics[width=3.25in]{fig/privacy.eps}
%\caption{The three categories of tasks and the authorized service providers that can perform the different task categories.}\label{fig:privacy_table}
%\end{figure}
%
%Privacy is a major concern in distributed processing of ``personal sensor data'' (e.g., vital-signs, location)  on geographically proximal volunteered computing resources.
Certain tasks in an application might work with sensitive (personal) user data that the user does not want to share. In such situations, these tasks can be given to only trusted service providers. To give different levels of protection to different tasks, we present a hierarchical approach similar to that of a social network. We assign tasks to different service providers based on level of trust, i.e., we determine what computing resources the different tasks are assigned to. We elaborate on our idea under the context of a ubiquitous health monitoring application. Data-analysis tasks are basic statistical methods that run over a tremendous amount of time-series data. The knowledge of the ``data type'' and context is inconsequential for the data-analysis tasks and, hence, the data can be anonymized so to not provide any private information (e.g., participant's identity and health status). Therefore, data-analysis tasks (\emph{public} tasks) can be performed on \emph{any} ``volunteered resource" in proximity without any concerns over the level of trust of the computing resource. %(i.e., any service provider)
Conversely, data-manipulation tasks (e.g., artifact removal in biomedical signals) need to be aware of the data type and, hence, can be carried out \emph{only} on ``trusted resources". However, they do not need any contextual information or identity of the participant whom the data belongs to. Trusted resources include service providers belonging to family members and friends (on social networks and real life) and this category of tasks is referred to as \emph{protected} tasks. Differently from the other two, decision-making tasks require the participant's identity and contextual information to generate baseline information (e.g., health status of participants in a biomedical application). Therefore, these \emph{private} tasks can \emph{only} be performed on the participant's ``personal mobile devices" (highest level of trust). %The aforementioned restrictions arising out of privacy concerns are taken into account in the task scheduling mechanism of \maestro\ in the form of constraints.
%Last, but not least, state-of-the-art data-encryption methods will be used to protect the patient's data moved over the wireless channel from eavesdropping and over-the-air manipulation (thus, maintaining integrity).

%Figure~\ref{fig:privacy_table} summarizes the three categories of tasks and shows the authorized service providers that can perform the different task categories based the task sensitivity.

\subsection{Task Deduplication}
%\textbf{Task Deduplication: }
To handle duplicate tasks in different workflows arriving at the broker, we present a task deduplication and workflow consolidation mechanism. The brokers group service requests before proceeding with task deduplication. The duration (time window) for which a broker waits ($\delta^{wait}_b~\mathrm{[s]}$) before deduplication is a tunable parameter. For a given rate of service request arrivals, the larger the window, the greater the chances of finding task duplicates. However, the windows cannot be too large as the workflow requests have to be serviced real time. This ``pause-aggregate-service'' strategy eliminates the unrealistic assumption of \emph{strictly simultaneous} workflow arrivals at the broker.
%The brokers receive multiple workflow execution requests over a period of time from the service requesters. The brokers group service requests and forward the workflow descriptions to their meta-broker from time to time. The meta-broker again aggregates requests from multiple arbitrators before it proceeds with task and workflow deduplication. The duration (time window) for which a broker waits ($\delta^{wait}_b$) before forwarding requests to a meta-broker and the duration for which the meta-broker waits ($\delta^{wait}_m$) before deduplication are tunable parameters. For a given rate of service request arrivals, the larger the windows the greater the chances of finding task and workflow duplicates. However, the windows cannot be too large as the workflow requests have to be serviced real time. This ``pause-aggregate-service'' strategy eliminates the simplifying assumption of strictly simultaneous workflow arrivals at the brokers and the meta-broker.
%What do arbitrators do? Aggregation of tasks arriving in a window eliminates jitter.

\dedup\----at the broker---parses the workflow descriptions to identify task duplicates and to create simplified workflows (with fewer tasks than before). \dedup\ looks for matching sub-graphs (connected group of tasks) between a pair of DAGs. Trivially, every single vertex in a DAG (workflow) is a sub-graph. \dedup\ starts with the comparison of Stage-0 tasks in the two workflows, as shown in Algorithm~\ref{algo:dedup}. Two tasks are considered to be ``similar'' when the following attributes match: task identifier (i.e., type), number and types of inputs (i.e., set of parent tasks, $\mathcal{P}$), and inputs' sizes (quantity of inputs), as shown in Algorithm~\ref{algo:checkSimilarity}. When tasks in two DAGs are similar, their corresponding sets of child tasks ($\mathcal{C}$s) are recursively checked for similarity. This recursive step is aimed at growing the size (i.e., number of tasks) of the matched sub-graph. In the recursive procedure, when the tasks under comparison, say task $k$ of workflow~1 and $l$ of workflow~2, cease to be similar, a link is created from $k$'s parent to $l$. Also, $l$ is added to the children set of Parent($k$). The tasks belonging to the duplicate subgraph in workflow~2 are discarded. Note that, while checking for similarity, tasks that have been visited and have tested positive for similarity are marked, so that they need not be checked again. The worst-case time complexity of \dedup\ is $\mathcal{O}(|\mathcal{V}_1|\cdot|\mathcal{V}_2|)$, where $\mathcal{V}_1$ and $\mathcal{V}_2$ are the sets of vertices in the two input DAGs.

%These attributes help identify the most appropriate ``implementation'' of a particular task (when multiple implementations exist) that would interface seamlessly with the parent and child tasks in the corresponding workflows.

\begin{algorithm}[t!]
\small
\caption{\dedup} \label{algo:dedup}
\textbf{Input:}  $\mathcal{K}$ and $\mathcal{L}$ are initially set to stage-0 tasks of the concurrent workflows \\
\textbf{Output:} Simplified workflows
%Set $\mathcal{A}^*$ of feasible solutions (associativity matrices $\mathbf{A}$s)
%\vspace{-2ex}
\begin{algorithmic}
\FOR{every $k$ $\in$ $\mathcal{K}$}
%\STATE \COMMENT {Construct the bi-adjacency matrix $\mathbf{C}^*$}
\IF{(visited($k$) == {\tt true})}
    \STATE continue
\ENDIF
\FOR{every $l$ $\in$ $\mathcal{L}$}
\IF{(visited($l$) == {\tt true})}
    \STATE continue
\ENDIF
\IF{({\tt checkSimilarity}($l$,$k$) == {\tt true})}
    \STATE{\dedup($\mathcal{C}^k$,$\mathcal{C}^l$)}
\ELSE
\STATE Parent($l$) = Parent($k$)
\STATE addChild(Parent($k$),$l$)
\ENDIF
\ENDFOR
\ENDFOR
\end{algorithmic}
\end{algorithm}
\begin{algorithm}[t!]
\small
\caption{{\tt checkSimilarity}} \label{algo:checkSimilarity}
\textbf{Input:}  Tasks $k$ and $l$ \\
\textbf{Output:} {\tt true} or {\tt false}
%\vspace{-2ex}
\begin{algorithmic}
\IF{($k.taskID$ == $l.taskID$) AND ($k.output$ == $l.output$)}
\IF{{\tt checkSimilarity}($\mathcal{P}^k$,$\mathcal{P}^l$)}
\STATE visited($k$) = visited($l$) = {\tt true}
\STATE visited($\mathcal{P}^k$) = visited($\mathcal{P}^l$) = {\tt true}
\RETURN {\tt true}
\ELSE
\RETURN {\tt false}
\ENDIF
\ELSE
\RETURN {\tt false}
\ENDIF
\end{algorithmic}
\end{algorithm}

\begin{figure*}[t!]
\centering
\begin{tabular}{ccc}
\includegraphics[width=2.12in]{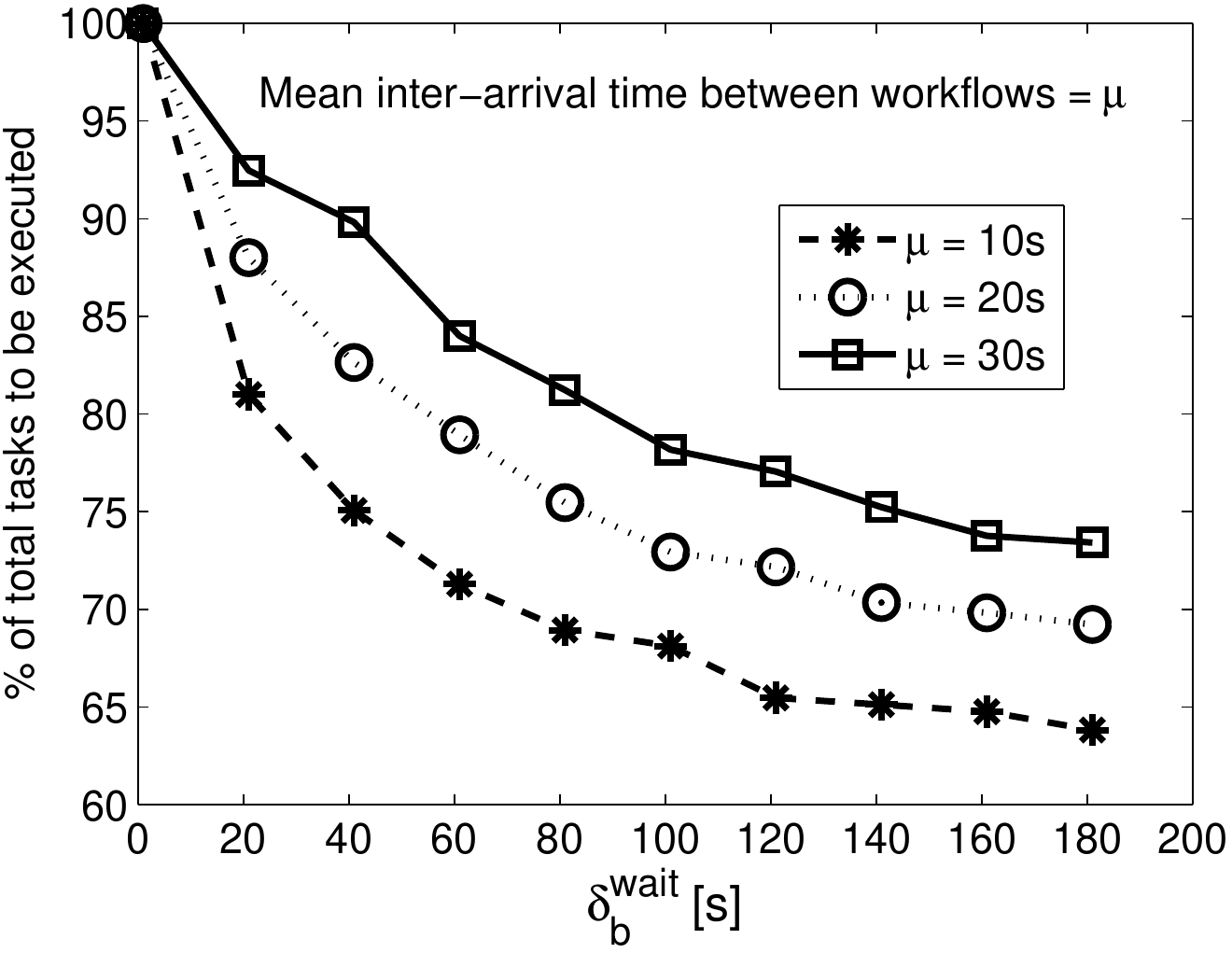} &
\includegraphics[width=2.12in]{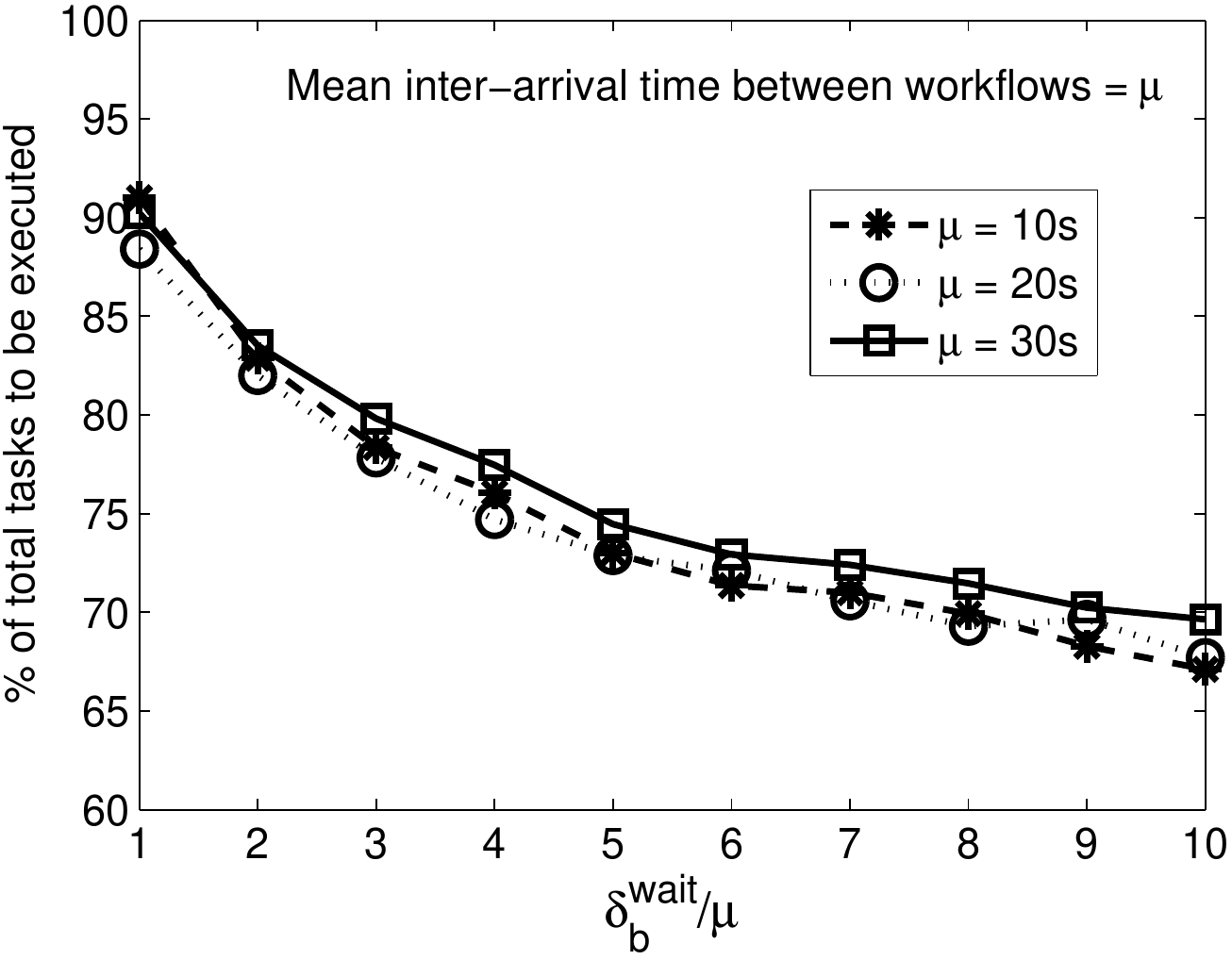} &
\includegraphics[width=2.12in]{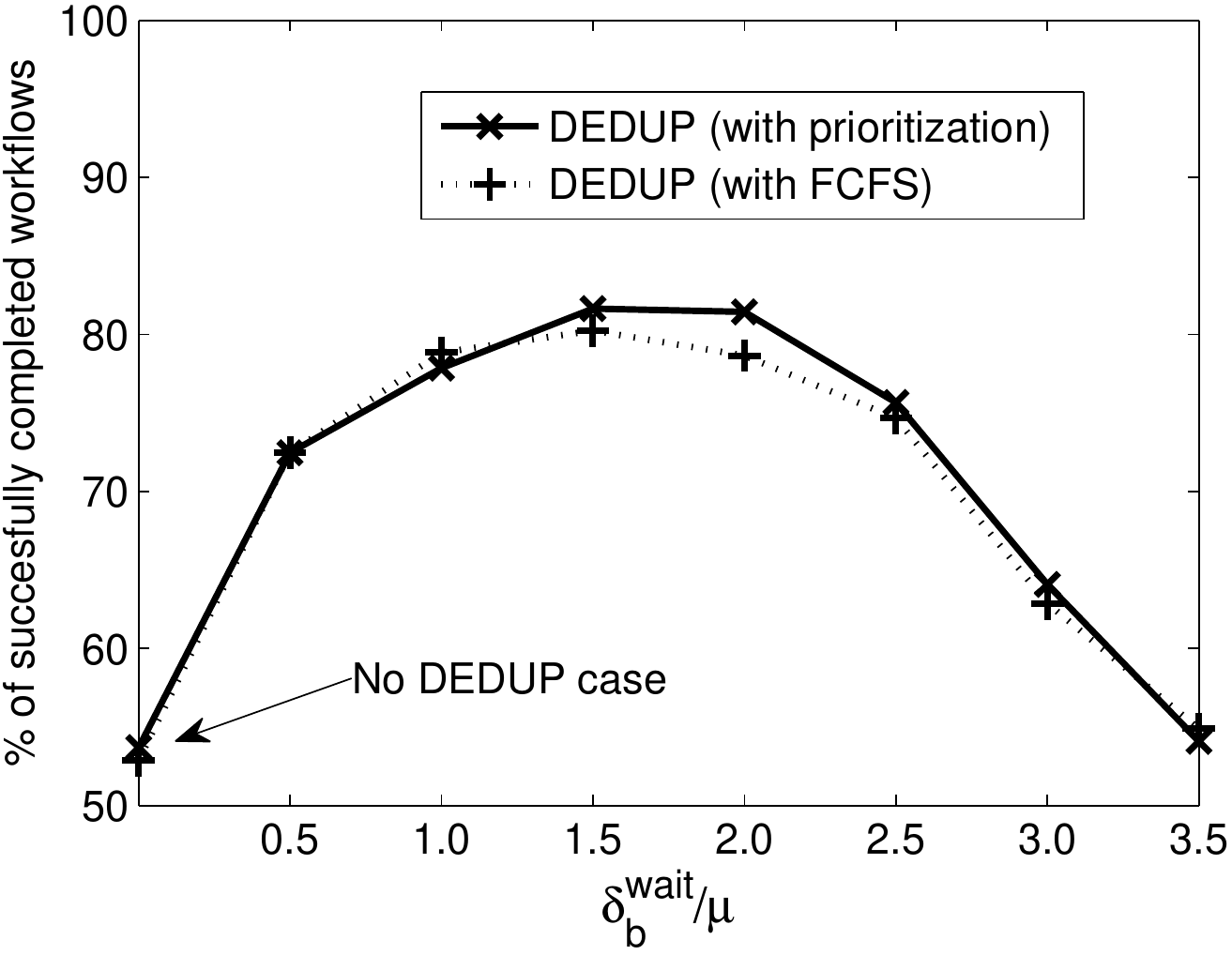} \\
\small (a) & \small (b) & \small (c)
\end{tabular}\caption{Decrease in percentage of the total number of tasks to be executed (while using \dedup) with increase in $\delta_b^{wait}$ (a) when $\delta_b^{wait}$ is not adapted to the arrival rate of workflows; (b) when $\delta_b^{wait}$ is adapted to the arrival rate of workflows (proportionally); (c) Behavior of \dedup\ in terms of percentage of successfully completed workflows with increase in $\delta_b^{wait}$.}\label{fig:dedup}
\end{figure*}

In the resulting workflow, $k$'s immediate predecessor task will be ``fork'' point from which other deduplicated workflow branches out. In Fig.~\ref{fig:applications}(a), the vertices corresponding to Peak detection in ECG signal and to Activity scoring of accelerometer output become fork points. Note that the time complexity of \dedup\ (or the total number of comparisons) is not altered by changing the order of comparison of different DAGs. Similarly, \dedup\ results in the same set of simplified workflows irrespective of the order of comparison of the different DAGs. It is not necessary that deduplication is done strictly before task allocation; deduplication is also achieved on the fly when tasks are already in execution. Under such circumstances, execution of duplicates is piggybacked, thus deduplicating at run time. Also, results of certain repetitive tasks are cached locally at the SPs so to deduplicate services. \emph{Service forwarding}, either at run time or through caching, relies on inferential analysis from historical request traces.

\section{Performance Evaluation}\label{sec:evaluation}
We developed a simulator in Java$^{\textsc{TM}}$ to evaluate empirically the performance gains provided by the different components of \maestro. Simulations also allow us to evaluate at scale. In the following, firstly, we present details about our experiment methodology, specifically, the workflows, workflow traces, the service providers, and the SP dynamics. Then, we discuss specific simulation scenarios as well as provide the results that (i) demonstrate the benefits of \dedup, (ii) highlight the price of using different protection levels in mobile computing, and (iii) illustrate the merits of replication-based failure handling.

%\subsection{Methodology}
%\textbf{Methodology: }
\textbf{Workflows: }The workflows we used, inspired by the applications discussed earlier, are all task parallel with data-parallel sub-graphs built into some of them. Even though currently available example workflows (from the biomedical and robotics domains) can be used for preliminary evaluation of \maestro, evaluation at scales necessitates the creation of arbitrary workflows that are similar to the ones available in literature. Hence, we developed a workflow (DAG) generator to generate \emph{arbitrary} workflows for large simulations. The synthetic workflows used in our simulations vary in terms of number of stages, tasks per stage, types and sizes of tasks at each stage and dependencies, and are representative of a wide range of task-parallel applications that \maestro\ can support.

\textbf{Workflow Traces: }At different instants, the service requesters (which may also be data providers) submit workflow requests to brokers while also specifying a deadline and a probability of success. The utility of the result from the workflow is assumed to be zero after the requester-specified deadline. Traces that capture workflow request arrivals over time in a mobile computing environment are not available in the literature. Workflow arrival traces in cloud and grid-computing environments cannot be adopted directly either. This is because the workflows, their deadline requirements, and arrival statistics are not representative of the applications or of the dynamics envisioned in \maestro. Hence, we developed a workflow-trace generator that can create multiple workflow arrival traces varying in terms of number of workflows and inter-arrival time between workflows as well as request-specific deadline and probability of success.
% or \rev{this last ``or" lost me...} any other work on MDCs and opportunistic computing
%For near real-time performance, the delay needs to be in the order of tens of seconds and the numbers clearly motivate the need to divide the tasks among service providers in the vicinity for speed up.

\begin{table*}[t!]
\caption{Characteristics of the heterogeneous mobile computing devices in our testbed.}\centering
\label{table:heterogeneity}
\begin{center}
\small
\begin{tabularx}{\linewidth}{|c|X|X|X|X|X|}
  \hline
    \shortstack{ \\ \\ \textbf{Devices}} & \textbf{Samsung Galaxy Tab} & \textbf{ZTE Avid N9120} & \textbf{Huawei M931} & \textbf{Toshiba Satellite} & \textbf{Raspberry Pi}\\
   \hline
   \hline
  {Type of devices} & Tablet & Smartphone & Smartphone & Laptop & Netbook\\
  \hline
  {No. of devices} & 2 & 3 & 1 & 1 & 1\\
  \hline
  \shortstack{ \\ \\ CPU} & 1GHz~Dual-core ARM & 1.2GHz~Dual-core & 1.5GHz~Dual-core & 2.13~GHz~i3~Intel  & 700~MHz ARM \\
  \hline
  {OS} & Android v4.0 & Android v4.0 & Android v4.0 & Windows 7 & Windows 7\\
  \hline
  \shortstack{ \\ \\ RAM~[GB]} & 1 & 0.512 & 1 & 4 & 0.512 \\
  \hline
  \shortstack{ \\ \\ Battery~[mAh]/[V]} & 7,000/4 & 1,730/5 &  1,650/10.8 & 4,200/10.8 &  2,200/5 \\
  \hline
%   \shortstack{ \\ \\ \textbf{Battery volt~[V]}} & 10.8  & 11.1 & 11.1 \\
%  \hline0
\end{tabularx}
\end{center}
\end{table*}

\begin{comment}
\begin{figure}[t!]
\centering
\includegraphics[width=3.3in]{testbed}
\caption{Architecture of \maestro\ including service requester and providers.}\label{fig:testbed}
\end{figure}
\end{comment}

\begin{table}[t!]
\caption{Average execution times of tasks of robotic applications on our testbed.}%\centering
\label{table:heterogeneity2}
\begin{center}
%\small
\begin{tabularx}{\linewidth}{|p{2.30cm}|X|X|X|}
  \hline
  \textbf{Computing Task} & \textbf{Samsung Galaxy Tab} & \textbf{Raspberry Pi}  & \textbf{Toshiba Satellite Laptop}\\
  \hline
    \hline
  Location determination~[s] & 143 & 1100 & 28.6  \\
  \hline
  Field estimation~[s] & 28.5 & 273  & 5.7 \\
  \hline
  Error estimation~[s] & 0.3 & 3.3 & 0.06  \\
  \hline
  Energy estimation~[s] & 0.3 & 3.3  & 0.06   \\
 \hline
\end{tabularx}
\end{center}
\end{table}

\textbf{Service Providers: }In our simulations we use a heterogeneous pool of SPs. The factors that contribute to heterogeneity are processing speed or capability (in terms of number of instructions per second), communication capability (in terms of $\mathrm{bps}$), rate of battery drain for computation (in terms of $\mathrm{mAh}$ per instruction), and finally the duration of availability. We use the mean availability duration (the duration for which the SP is in the MDC) and the mean away duration (the duration for which the SP is not in the MDC) as well as their respective distributions to control the dynamics in the mobile computing environment. We choose these durations carefully to maintain an average number of SPs in the MDC as the three variables are related by Little's theorem.

%\subsection{Benefits of Task Deduplication}
\textbf{Benefits of Task Deduplication: }To demonstrate the benefits of task deduplication, we performed two experiments to ascertain the following two factors with and without \dedup: (i) the reduction in percentage of total number of tasks to be executed in the MDC and (ii) the percentage of successfully completed workflows among all the requests submitted.

\textbf{Experiment~1: }As $\delta^{wait}_b$ is a tunable parameter, we observed performance in terms of reduction in number of tasks by varying it. We studied the behavior of \dedup\ when $\delta^{wait}_b$ is adapted to the arrival rate of workflows and when it is not. We created three distinct workflow traces, each with $100$ requests, with mean inter-workflow-arrival durations of $\mu=10,20,$ and $30~\mathrm{s}$, respectively. The number of distinct workflows in each trace was set to $40$, and the number of SPs to $10$.%The deadline of each workflow request is set randomly between $40$ and $80s$.

\emph{\underline{Observations}: }Figures~\ref{fig:dedup}(a) and \ref{fig:dedup}(b) show that the percentage of total number of tasks to be executed in the MDC decreases by $25\%$ when $\delta_b^{wait}$ is five times the inter-arrival duration $\mu$ of the workflows. This decrease will be greater when the number of distinct workflows in the traces is reduced below the current value of $40$. Figure~\ref{fig:dedup}(a) was obtained by varying $\delta_b^{wait}$ in increments of $20$ agnostically to the workflow arrival rate. As a result, in comparison to the trace with $\mu=20s$, the percentage of tasks to be executed is higher for the trace with $\mu=30~\mathrm{s}$ while it is lower for the one with $\mu=10~\mathrm{s}$. This is because for the same $\delta_b^{wait}$, the number of workflows considered together for deduplication decreases with increase in $\mu$. Therefore, adaptation of $\delta_b^{wait}$ with respect to $\mu$ is key to achieve improved performance (see Fig.~\ref{fig:dedup}(b)).

\textbf{Experiment~2: }We observed the performance in terms of percentage of successful workflow completions by varying the waiting period $\delta^{wait}_b$. We created a workflow trace with a total of $500$ requests. The mean inter-workflow-arrival duration in the workflow trace was set to $\mu=10~\mathrm{s}$ and the number of distinct workflows in the trace was set to $10$. The deadline of each workflow request was chosen randomly between $40$ and $80~\mathrm{s}$, and the number of SPs was set to $10$.

\emph{\underline{Observations}: }Figures~\ref{fig:dedup}(c) shows that the percentage of successful workflow completions in the MDC increases to as much as $83\%$ (compared to the baseline no \dedup\ case at $53\%$) when $\delta_b^{wait}$ is twice the inter-arrival duration $\mu$ of the workflows. This increase will be greater when the failed workflow tasks are discarded (which we did not do to study the worst case). Figure~\ref{fig:dedup}(c) clearly highlights the situation when \dedup\ may not be beneficial in \maestro. Even though an increase in $\delta_b^{wait}$ results in a decrease in the total number of tasks to be executed, as shown in Figs.~\ref{fig:dedup}(a) and \ref{fig:dedup}(b), the decrease is only sub-linear. The accumulation of tasks over time may result in an overload for the underlying SP pool as is the case when $\delta_b^{wait}/\mu > 2.0$ in Fig.~\ref{fig:dedup}(c) where the gain drops until finally reaching the baseline at $\delta_b^{wait}/\mu = 3.5$. Also, it is important to note that the task prioritization in \maestro\ results in improved performance in comparison to a First-Come-First-Served~(FCFS) scheduling policy. The difference in performance will widen further when the variance in deadlines is greater than what we used here. %What was infeasible was made feasible by DEDUP XX. Initial 50\%. Mention baseline case od ASAP processing.

\begin{comment}
\begin{figure}[h!]
\centering
\includegraphics[width=3.3in]{protection_gain}
\caption{Percentage of successful workflow completions under different SP dynamics in the MDC. Scenarios A through E represent a progressive decrease in the stability of SPs. $\widetilde{\lambda}^{-1}~\mathrm{[s]}$ is the average inter-arrival duration and $\widetilde{T}~\mathrm{[s]}$ is the average availability duration of the SPs.}\label{fig:protection}
\end{figure}
\end{comment}

%\subsection{Benefits of Controlled Replication}

\begin{figure*}[t!]
\centering
\begin{tabular}{ccc}
\includegraphics[width=2.1in]{testbed} &
\includegraphics[width=2.00in]{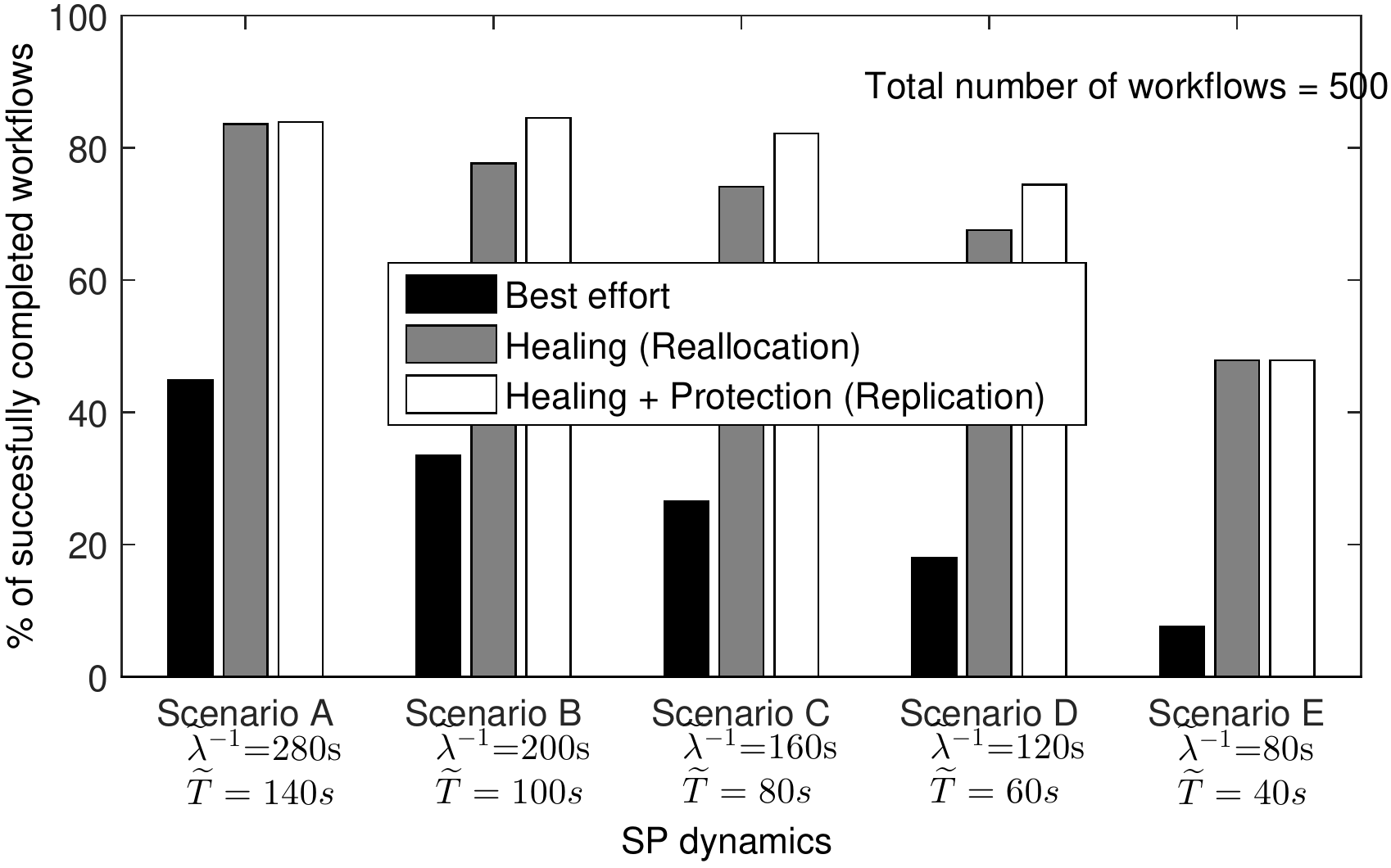} &
\includegraphics[width=2.00in]{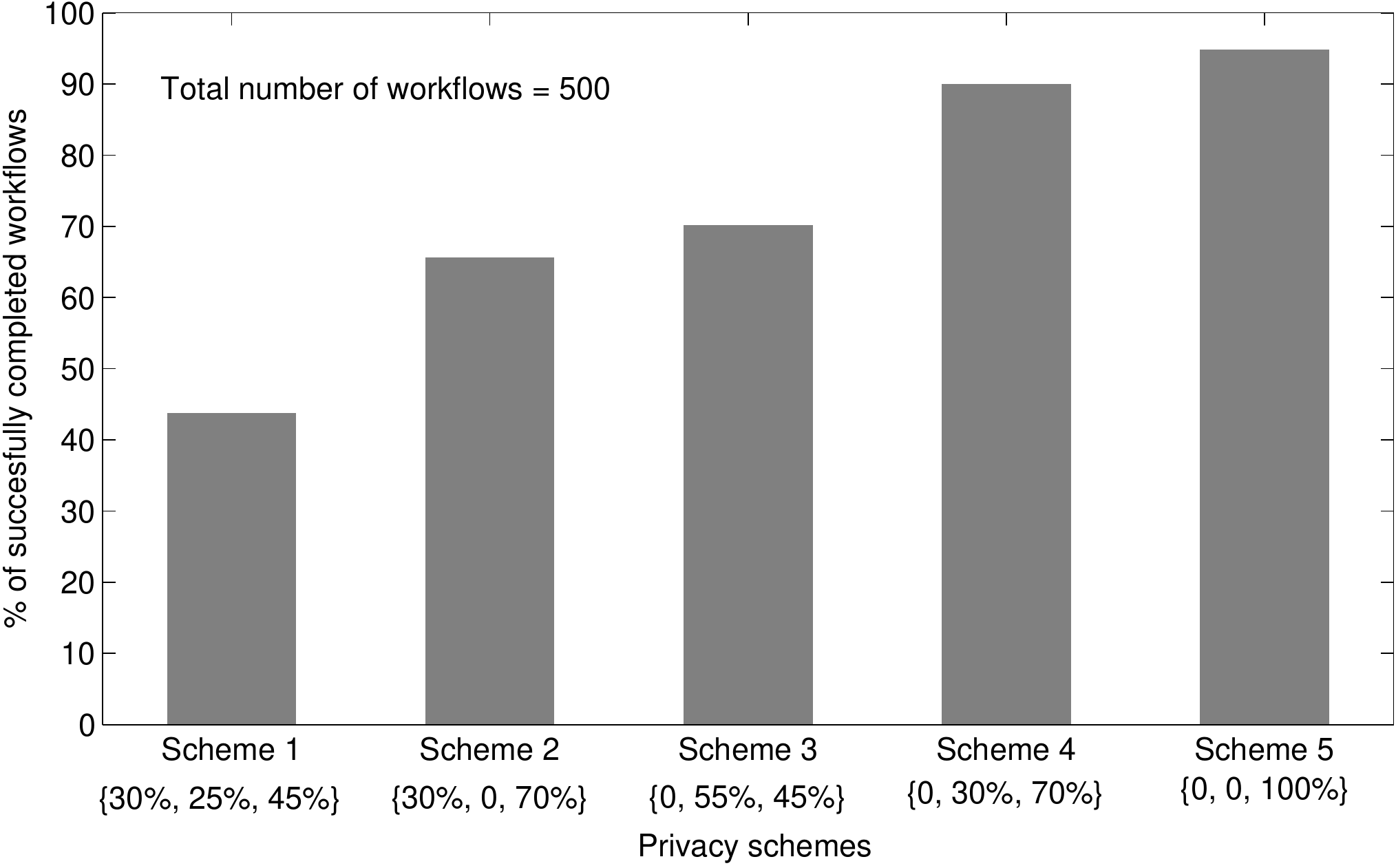} \\
\small (a) & \small (b) & \small (c)
\end{tabular}\caption{(a) Architecture of \maestro\ including service requester and providers; (b) Percentage of successful workflow completions under different SP dynamics in the MDC. $\widetilde{\lambda}^{-1}$ [s] is the average inter-arrival duration and $\widetilde{T}~\mathrm{[s]}$ is the average availability duration of the SPs; (c) Percentage of successful workflow completed when different levels of protection schemes are employed. }\label{fig:PE_2}
\end{figure*}

%\subsection{Testbed}
\textbf{Testbed: }The focus of this subsection is geared towards presenting the performance of our solution for real-world workflows. We prepared a testbed for our experiments, which consisted of Android- and Linux-based mobile devices with heterogeneous capabilities (summarized in Table~\ref{table:heterogeneity}). We consider biomedical, robotic, and computer vision workflows presented in Fig.~\ref{fig:applications} to show the performance of our proposed controlled replication approach in order to overcome uncertainty due to failure of SPs to finish allocated task. We profiled the time taken for various tasks in the workflow on all the devices in our testbed. The architecture of our testbed is shown in Fig.~\ref{fig:PE_2}(a), which is based on our work in~\cite{Viswanathan12_TPDS}. The service requester device contains the resource task mapper, which is responsible to allocate task to different service providers.

\emph{\underline{Input data set}: }For each of application given in Fig.~\ref{fig:applications} we obtained the execution time of each task by extensive offline profiling which involved running each task of an application with multiple input data. For applications in Fig.~\ref{fig:applications}(a,b) we generated the input data artificially and for application in Fig.~\ref{fig:applications}(c) we used input data as images from the Berkeley image segmentation and benchmark dataset~\cite{MartinFTM01}.

\textbf{Motivation for Controlled Replication: }We utilize real-time distributed robotic applications, as in Fig.~\ref{fig:applications}(b), to motivate the need for controlled replication. For such applications, e.g., detection of harmful chemicals in a field, quick execution of tasks given in the workflow is essential. Table~\ref{table:heterogeneity2} shows the set of tasks in the robotic application and execution times of these tasks. We see that location-determination and field-estimation tasks are critical because of their high execution times. If we fail to get results for these tasks from a SP due to network disconnection or lack of resources at that SP, the workflow may not meet the deadline. To avoid unnecessary idle waiting times incurred in reactive failure handling, a proactive approach like our proposed controlled replication will help meet the application deadline.

\textbf{Benefits of Controlled Replication: }\maestro's task-scheduling mechanism employs proactive protection (selective controlled replication of large tasks) in addition to reactive healing (reallocation of failed tasks) in order to provide robustness. To study the improvement in performance provided by healing  and protection over the best-effort (baseline) case, we performed an experiment under different service provider dynamics. We set the \emph{average number of active SPs} in the MDC to $30$. We varied the SP dynamics in the MDC from highly volatile to highly stable by tuning the following parameters---\emph{average inter-arrival duration} ($\widetilde{\lambda}^{-1}~\mathrm{[s]}$) of SPs and \emph{average availability duration} ($\widetilde{T}~\mathrm{[s]}$) of SPs. The average number of SPs and the aforementioned parameters are related by Little's law. We created a workflow trace with a total of $500$ requests. The mean inter-workflow-arrival duration in the trace was set to $\mu=20~\mathrm{s}$ and the number of distinct workflows was set to $10$. The trace has a mix of small ($66\%$) and large ($33\%$) workflows (differing in terms of task sizes and deadlines). The deadline of the small workflows was chosen randomly between $40$ and $80~\mathrm{s}$, while the large workflows' deadline was picked randomly between $80$ and $160~\mathrm{s}$. 

\emph{\underline{Observations}: }Figure~\ref{fig:PE_2}(b) shows five ordered scenarios, A through E, where A corresponds to a highly stable MDC and E corresponds to a highly volatile one. The performance of \maestro\ with only healing as well as with both healing and protection is always better than the baseline case. In Scenario~B through D the use of protection in addition to healing prevents more workflows from failing than using only plain self-healing. This is because when using healing in isolation the time taken to recover from a failure of a task belonging to a ``large'' workflow is more than twice that the task execution time. However, selective replication of such critical tasks (which may easily jeopardize the workflow when they fail), as done in protection, prevents a greater percentage of workflows from failing. However, note that protection does not provide any additional gain over plain healing in Scenario~A when the probability of SP failure during task execution is very low and in Scenario~E when the probability is very high.

\begin{comment}
\begin{figure}[b!]
\vspace{-0.15in}
\centering
   \begin{subfigure}{0.45\textwidth}
   \centering
   \includegraphics[width=0.9\linewidth]{fig/PerfSPSH_time.eps}
   \caption{}
  % \label{fig:Ng1}
\end{subfigure}
\begin{subfigure}{0.45\textwidth}
%\vspace{-0.1in}
\centering
   \includegraphics[width=0.9\linewidth]{fig/PerfSPSH_wf.eps}
   \caption{}
   %\label{fig:Ng2}
\end{subfigure}
%\vspace{-0.2in}
\caption{Comparison of performance of controlled replication and self-healing in terms of (a) Execution time by varying the SP statistics in the MDC; (b) Percentage of successful workflows by varying the task sizes.}
\label{CompareSPandSH}
\vspace{-0.25in}
\end{figure}
\end{comment}

%\subsection{Price of Using Multiple Protection Levels}
\textbf{Price of Using Multiple Protection Levels: }Even though \maestro\ provides multiple levels of protection to different tasks through controlled access by authorized service providers, there is a price to pay for it in terms of performance. Authorizing SPs to execute only certain types of tasks restricts the feasibility region of the solution to the problem the broker is trying to solve. In \maestro, the broker aims at scheduling tasks in the MDC in such a way that the percentage of successfully completed workflows be maximized. We performed an experiment to quantify the difference in performance when multiple levels of protection are employed by varying the percentage of private, protected, and public tasks in the workflows while keeping the percentage of personal, trusted, and untrusted 3$^{rd}$ party devices in the MDC fixed. These percentages are $1$, $33$, and $66\%$, respectively. We created a workflow trace with a total of $500$ requests. The mean inter-workflow-arrival duration in the workflow trace was set to $\mu=10~\mathrm{s}$ and the number of distinct workflows in the trace was set to $10$. The deadline of each workflow request was chosen randomly between $40$ and $80~\mathrm{s}$. %The \cumulus\ has 10 SPs.

\emph{\underline{Observations}: }Figure~\ref{fig:PE_2}(c) shows the five schemes with decreasing degree of levels of protection (or increasing number of public tasks in the workflows). We observed that, as the degree of protection is decreased, i.e., the percentage of public tasks increased, the performance in terms of percentage of workflow completions increased. Scheme~5 represents an unrestricted scenario where all tasks are public, while Scheme~1 is extremely restricted. Scheme~2 through 4 reflect what may be adopted in real-world deployments. The performance of Schemes~2 and 3 can be improved to match that of Scheme~4's by either relaxing the deadline requirements or by increasing the MDC size. A small relaxation in the deadline is a marginal cost to incur if multiple protection levels are desired.
\begin{comment}
\begin{figure}[t!]
\centering
\includegraphics[width=3.3in]{fig/privacy_schemes.eps}
\caption{Percentage of successful workflow completions when different levels of protection are employed. The tuple represents the \% of private, public, and protected tasks in each scheme. Schemes~1 through 5 represent a progressive decrease in the degree of protection.}\label{fig:privacy_price}
\end{figure}
\end{comment}

\begin{figure*}[t!]
\centering
\begin{tabular}{ccc}
\includegraphics[width=2.10in]{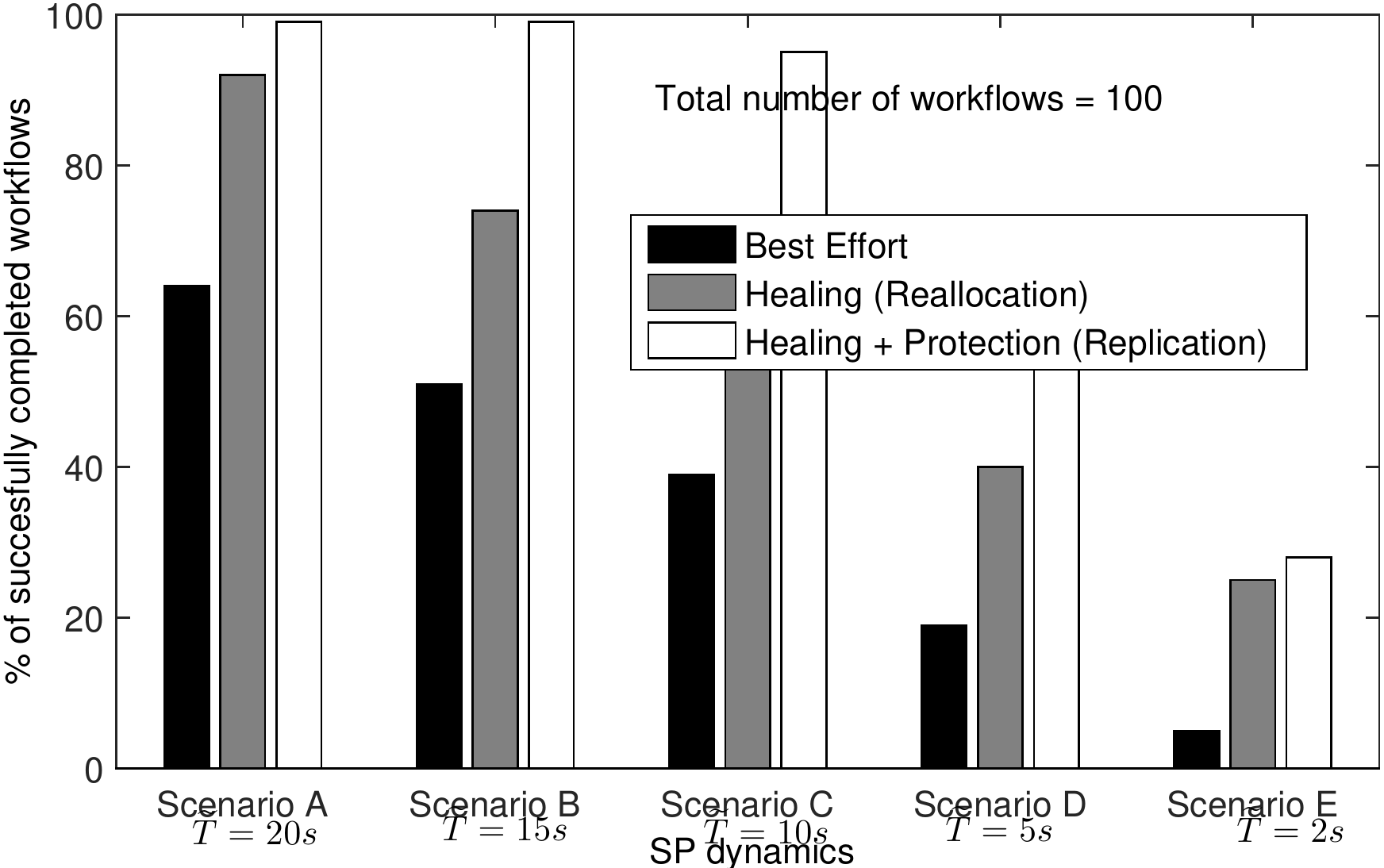} &
\includegraphics[width=1.95in]{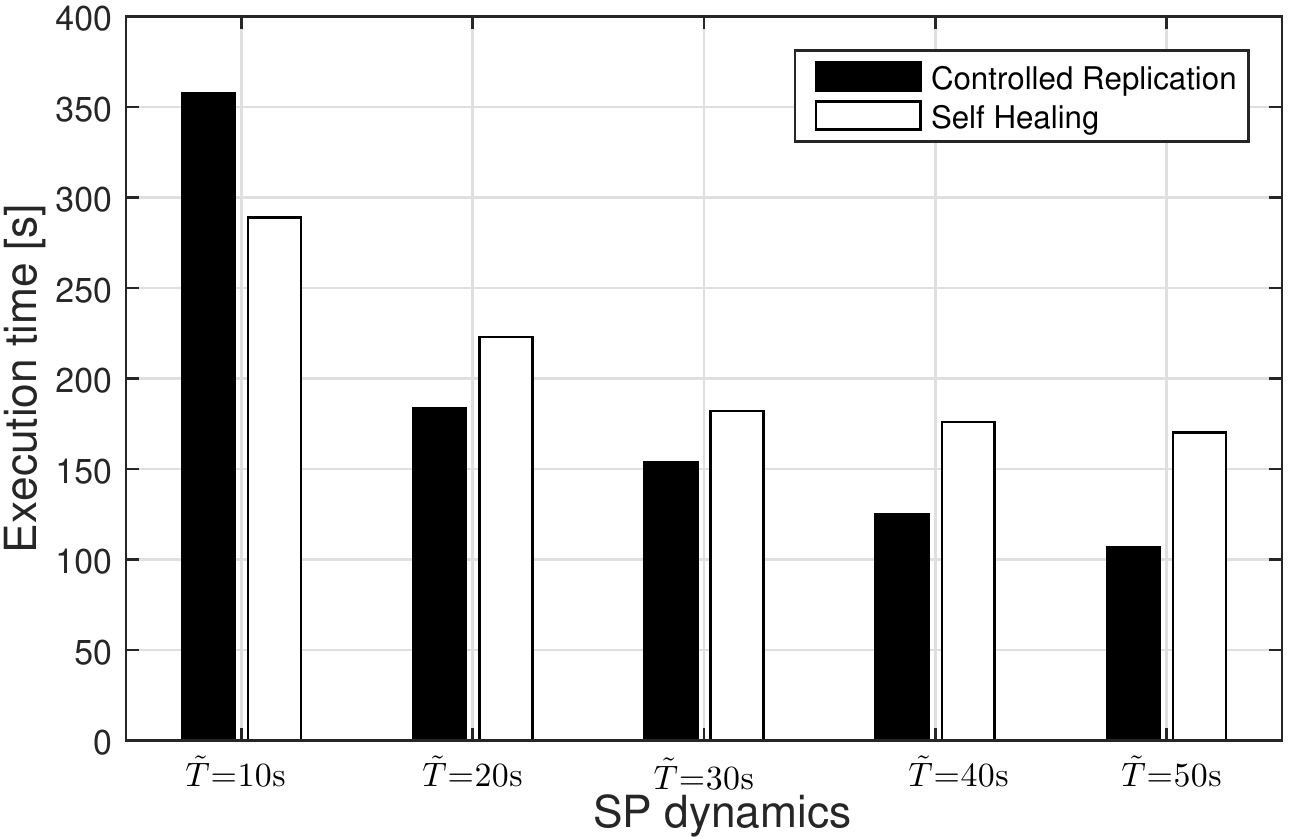}&
\includegraphics[width=1.95in]{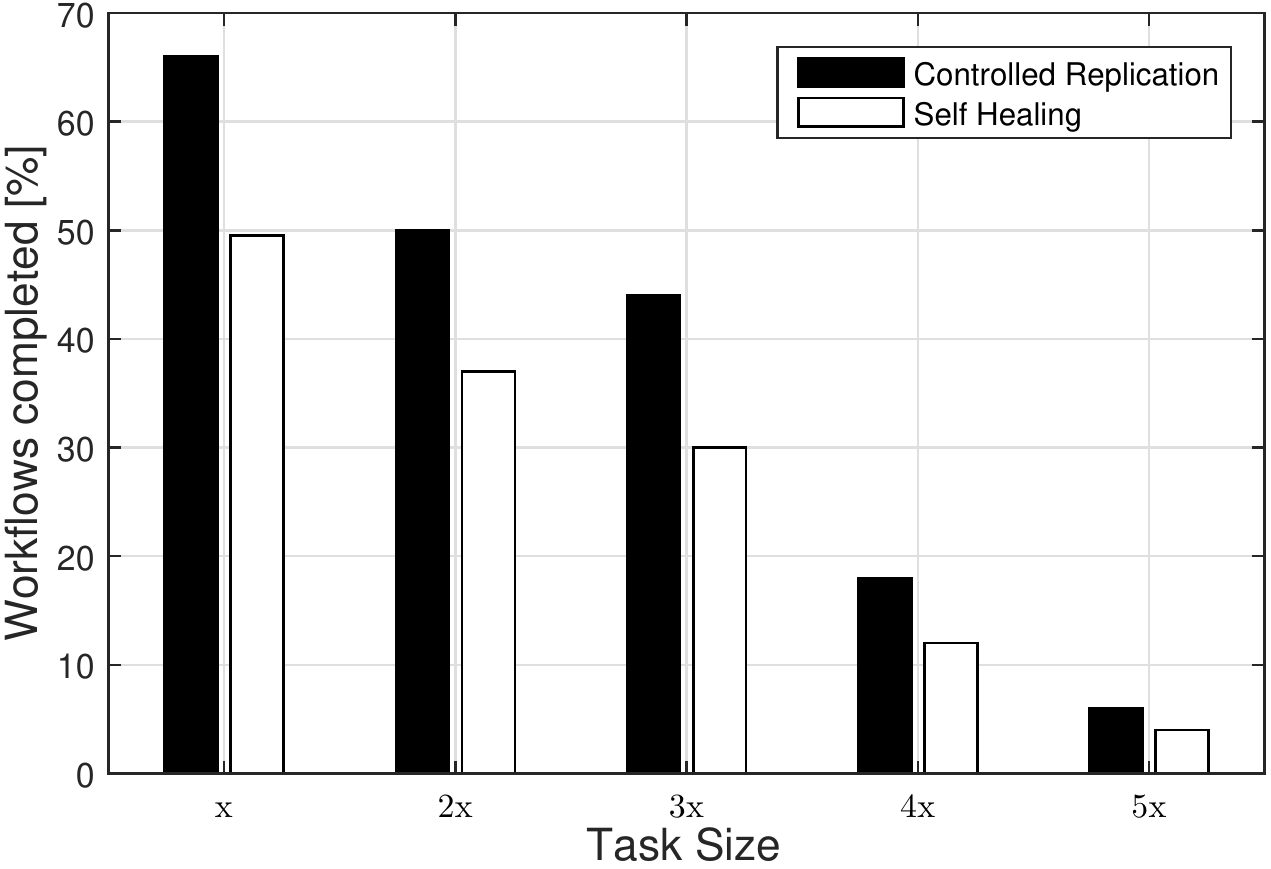} \\
\small (a) & \small (b) & \small (c)
\end{tabular}
\caption{(a) Comparison of controlled replication technique with best effort and healing approaches via experiments on our testbed; Comparison of performance of controlled replication and self-healing in terms of (b) execution time by varying the SP statistics in the MDC; (c) percentage of successful workflows by varying the task sizes. }\label{CompareSPandSH}
\end{figure*}

\textbf{Controlled Replication for Biomedical Applications:}
We focus on the stress-detection application, which receives input data from variety of smartphone sensors. To study the improvement in performance provided by self-healing and self-protection over the best-effort (baseline) case, we performed an experiment under different service-provider dynamics. The goal of our technique protection is controlled replication of collective tasks in the workflow. We studied the performance of our approach in case of real-time deadline constraints. From the concepts explained earlier, we determined the anxiety-detection task to be the most critical task of the workflow. As a result, this task was replicated on a subset of SPs based on~\eqref{eq:protec}.

\begin{comment}
\begin{figure}[h!]
\centering
\includegraphics[width=3.3in]{fig/exp_fig.eps}
\caption{Comparison of controlled replication technique with best effort and healing approaches via experiments on our testbed.}\label{fig:PE_2}
\end{figure}
\end{comment}

%\end{comment}

\emph{\underline{Observations}:}
Figure~\ref{CompareSPandSH}(a) shows five ordered scenarios, A through E, where A corresponds to a highly stable MDC and E to a highly volatile one. A successful workflow is the one that completes all its tasks within the user-specified deadline. As seen earlier, the performance of \maestro\ with only healing as well as with both healing and protection is always better than the baseline case. As the mean duration of service time becomes higher, we observe both healing and protection with healing to give similar performance. We see that the performance of baseline technique with $63\%$ successful workflows is much lower than protection with healing and only healing approach with above $85\%$ successful workflows. In some cases, healing performs worse than combination of healing and protection as healing is a reactive approach and leads to increase in makespan due to failure of certain tasks. These experiments show the robustness of our technique for real-world, real-time workflows.

\textbf{Controlled Replication for Computer-vision Applications:}
To study controlled replication in computer vision domain we implemented Workflow~c1.a as shown in Fig~\ref{fig:applications}(c). We consider two competing fault-tolerance mechanism, self-healing and controlled-replication. We consider one of the device in the testbed, (Samsung Galaxy Tab) to serve as the broker and it sends multiple execution requests of Workflow c.1 to two other devices (ZTE Avid, Huawei) forming the MDC. To implement self-protection, tasks given to device ZTE Avid, Huawei is also replicated on another devices from the testbed given in the Table~\ref{table:heterogeneity}.

The performance of both controlled replication and self-protection over different SP dynamics in the MDC is shown in Fig.~\ref{CompareSPandSH}(b). We observe in Fig.~\ref{CompareSPandSH}(b) that for mean availability duration, $\tilde{T}=\mathrm{10s}$, self-healing performs better than self-protection, however, as the mean availability duration of the SPs increases controlled-replication performs better. However, this gain comes at the cost of employing higher the number of SPs than in the self-healing approach. We also calculate the percentage of successfully completed workflows within $\mathrm{200s}$ for different task sizes in the workflow given in Fig.~\ref{CompareSPandSH}(c). The  mean availability duration of SPs is considered to be, $\tilde{T}=\mathrm{40s}$. The task size here corresponds to the input data size $k^0_1$ in Workflow~c.1. As discussed earlier that as the task size increases reactive approach such as self-healing does not perform well as it leads to wastage of resources and time in case of failure of execution of task at the SPs. We observe that via controlled replication we are able to execute higher number of workflows in the same amount of time. As the task sizes increase further we see that neither self-healing or controlled replication perform well. To improve the performance of MDC the availability duration of SPs should be increased which will increase the number of completed workflows.  

\section{Conclusion}\label{sec:conclusions}
%\todo{Call this section discussion. Do not have a huge summary. Have a simple 2 liner followed by a discussion on multiple broker scenarios. That would be a gradual transition to a note on future work.}
We presented our vision for harnessing the sensing, computing, and storage capabilities of mobile and fixed devices in the field \emph{as well as} those of computing and storage servers in remote datacenters to form a mobile/fixed device cloud. We discussed \maestro, a framework for robust and secure pervasive mobile computing in a mobile/fixed device cloud, which can enable a wide range of mobile applications that rely on real-time, \emph{in-situ} processing of data generated in the field. We focused on two key components of \maestro\----(i) a task scheduling mechanism that employs controlled task replication in addition to task reallocation for robustness and (ii) \dedup\ for task deduplication among concurrent workflows. A flexible architecture to impart multiple levels of protection to tasks is also discussed. We are currently working towards addressing security issues that arise when malicious nodes provide incorrect results and not just denial-of-service attacks. %We are also continuing validation using a heterogeneous mix of mobile application workflows differing in terms of communication-to-computation-cost ratios and degree of concurrency.
%
%
%We advocate the use of multiple brokers in \maestro\ to avoid a single point of failure. The brokers play a very important role in handling uncertainties caused by the unavailability of service providers or by the inaccuracy of task-completion-time estimates. In order to ensure that the unavailability of a broker (due to poor connectivity or hardware failure) does not lead to the failure of the entire system, each broker shares with all of its active data and service providers a \textit{list of alternate brokers} ranked according to their proximity (primary key) and physical addresses (secondary key). We advocate the use of a distributed self-election mechanism~\cite{Eun11} for assigning the appropriate number of brokers. The brokers receive multiple workflow-execution requests over a period of time from the service requesters. The brokers group service requests and forward the workflow descriptions to their \emph{meta-broker} (additional logical role played by one of the brokers) from time to time. The meta-broker again aggregates requests from multiple brokers before it proceeds with task and workflow deduplication. As future work, we will extend \maestro\ to include multiple-broker scenario, which is not a trivial extension but requires a more elaborate treatment.

%\textbf{Acknowledgment: }This work was supported by the US ONR-Young Investigator Program~(YIP) Grant No.~11028418.

%\balance
%\begin{small}
\bibliographystyle{IEEEtran}\scriptsize
\bibliography{ApproxComp,biblio_Mob_Cloud_v2,biblio_Mob_Cloud_v2}
%\end{small}

\vspace{-1cm}
\normalsize

\end{document}